\documentclass[a4paper,12pt]{article}
\usepackage{axodraw4j}
\usepackage{color}
\usepackage{pstricks}
\usepackage{amssymb,amsmath,epsfig,graphicx,psfrag}
\usepackage{cite}

\numberwithin{equation}{section}

\newcommand{\per}{\,}
\newcommand{\pp}[2]{p_{#1}\hspace{-0.ex} p_{#2}}
\newcommand{\pq}[2]{p_{#1}\hspace{-0.ex} q_{#2}}
\newcommand{\qq}[2]{q_{#1}\hspace{-0.ex} q_{#2}}
\newcommand{\sq}{\bigg[} 
\newcommand{\dq}{\bigg]} 
\newcommand{\sg}{\bigg\{} 
\newcommand{\dg}{\bigg\}} 
\newcommand{\pst}{\Big(} 
\newcommand{\pdt}{\Big)} 
\newcommand\eqcs{\raisebox{-2mm}{\rlap{\,{\rm cs}\,}} \raisebox{.0ex}{$\,=\,$}}

\newcommand{\ab}{\mathrm{ab}}              
\newcommand{\as}{\alpha_{\mathrm{S}}}
\newcommand{\bra}[1]{\langle #1 |}

\newcommand{\bs}{\boldsymbol}

\newcommand{\gab}{\gamma^\mathrm{(ab)}}     
\newcommand{\gna}{\gamma^\mathrm{(na)}}     
\newcommand{\J}{\boldsymbol{J}}             
\newcommand{\ket}[1]{| #1 \rangle}

\newcommand{\lra}{\leftrightarrow}
\newcommand{\M}{{\cal M}}               
\newcommand{\na}{\mathrm{na}}              

\newcommand{\pol}{\varepsilon}
\newcommand{\qb}{\bar{q}}

\newcommand{\cS}{{\cal S}}               

\newcommand{\sla}[1]{#1\hspace{-0.45em}/\hspace{0.em}}
\newcommand{\sy}[2]{\bigl(#1\, #2\bigr)_{sym}}
\newcommand{\T}{\boldsymbol{T}}                   
\newcommand{\tri}[1]{{\cal T}^{(d)}_{#1}}   
\newcommand{\ui}{\mathrm{i}}             
\newcommand{\ub}{\bar{u}          }       

\def\lra{\leftrightarrow}

\newcommand\g{g_{\mathrm{S}}}

\newcommand\gq{g}

\def\ep{\epsilon}

\def\beq{\begin{equation}}
\def\eeq{\end{equation}}
\def\beeq{\begin{eqnarray}}
\def\eeeq{\end{eqnarray}}
\def\cm{{\cal M}}
\def\bom#1{{\mbox{\boldmath $#1$}}}
\def\to{\rightarrow}

\newcommand{\la}{\langle}
\newcommand{\ra}{\rangle}

\def\nn{\nonumber}

\def\ID{1 \kern -.45 em 1}

\def\ket#1{|{#1}\ra}
\def\bra#1{\la{#1}|}

\def\ubar{{\overline u}}

\def\cbet0{b_0}

\def\bj{{\bom J}}
\def\bt{{\bom T}}
\def\btq{{\bom t}}

\def\w0abc{w_{\{ABC\}}}

\begin{document}
\begin{titlepage}
\renewcommand{\thefootnote}{\fnsymbol{footnote}}
\begin{flushright}
FTUV-22-1017.4807
\end{flushright}
\par \vspace{10mm}
\begin{center}
{\Large \bf Soft gluon-quark-antiquark emission\\ in QCD hard scattering
}
\end{center}

\par \vspace{2mm}
\begin{center}
{\bf Stefano Catani}~$^{(a)}$, {\bf Leandro Cieri}~$^{(b)}$, {\bf Dimitri Colferai}~$^{(a)}$\\
and {\bf Francesco Coradeschi}~$^{(c)}$

\vspace{5mm}

${}^{(a)}$INFN, Sezione di Firenze and
Dipartimento di Fisica e Astronomia,\\ 
Universit\`a
di Firenze,
I-50019 Sesto Fiorentino, 
Florence, Italy \\
\vspace*{2mm}
${}^{(b)}$Instituto de F\'{\i}sica Corpuscular, Universitat de Val\`{e}ncia --
Consejo Superior de Investigaciones Cient\'{\i}ficas, Parc Cient\'{\i}fic,
E-46980 Paterna, Valencia, Spain \\
\vspace*{2mm}
${}^{(c)}$OVI-CNR, I-50141 Firenze, Italy.

\vspace{5mm}

\end{center}

\par \vspace{2mm}
\begin{center} {\large \bf Abstract} \end{center}
\begin{quote}
\pretolerance 10000

We consider the radiation of a soft gluon ($g$) and a soft 
quark-antiquark ($q{\bar q}$) pair in QCD hard scattering.
In the soft limit the scattering amplitude has a singular behaviour that is factorized
and controlled by a soft current, 
which has a process-independent structure in colour space.
We evaluate the soft $gq{\bar q}$ current at the tree level for an arbitrary multiparton
scattering process. The irreducible correlation component of the current includes strictly non-abelian terms and also terms with an abelian character. 
Analogous abelian correlations appear for soft photon-lepton-antilepton emission in QED.
The squared current for soft $gq{\bar q}$ emission produces colour dipole and colour tripole
interactions between the hard-scattering partons. The colour tripole interactions are odd
under charge conjugation and lead to charge asymmetry effects. We consider the specific applications to processes with two and three hard partons, and we discuss the structure of
the corresponding charge asymmetry contributions. We also generalize our QCD results to the cases of QED and mixed QCD$\times$QED radiative corrections.

\end{quote}

\vspace*{\fill}
\begin{flushleft}
October 2022 
\end{flushleft}
\end{titlepage}

\vskip 1cm


\setcounter{footnote}{2}

\section{Introduction\label{s:intro}}

The large amount of high-precision data already taken at the CERN large hadron collider (LHC) demands theoretical predictions with a corresponding high precision. This situation will be further accentuated with the Run 3, which already started in the spring of 2022.

In the context of QCD, the theoretical accuracy is typically increased by performing
perturbative calculations of radiative corrections at higher orders in the 
strong coupling $\as$. The present high-precision frontier is represented by computations at the next-to-next-to-next-to-leading order (N$^3$LO) in the QCD coupling
(see, e.g., Ref.~\cite{Heinrich:2020ybq} and references therein).

In theories with massless particles, like QCD, scattering amplitudes lead
to infrared (IR) divergent contributions, and finite results are obtained by combining real and virtual radiative corrections in computations of physical observables.
The basic property that produces the cancellation of the IR divergences is their 
universal (i.e., process-independent) structure. The IR singular behaviour of the scattering amplitudes is indeed controlled by universal factorization formulae and by corresponding singular factors for emission of soft and collinear radiation
(see, e.g., Ref.~\cite{Agarwal:2021ais} and references therein).
The knowledge of these factorization formulae in explicit form is therefore very important
to practically organize and greatly simplify the cancellation of the IR divergences in 
perturbative calculations at various perturbative orders.

The cancellation mechanism of the IR divergences produces residual logarithmic contributions that are quantitatively large for a wide class of physical observables which are evaluated in kinematical regions close to the exclusive boundary of the phase space. These large contributions have to be computed at high perturbative orders, and possibly resummed to
all orders in perturbation theory (see, e.g., Refs.~\cite{Becher:2014oda, Luisoni:2015xha} and references therein). For instance, QCD resummation for transverse-momentum distributions has reached the next-to-next-to-next-to-leading logarithmic (N$^3$LL) accuracy
\cite{Luo:2019szz, Ebert:2020yqt, Luo:2020epw, Ebert:2020qef}. In general, soft and collinear factorization formulae of scattering amplitudes are important ingredients in the context of QCD computations and resummations of large logarithmic contributions of IR origin.



Soft and collinear factorization formulae at ${\cal O}(\as)$ had a key role to devise  process-independent and observable-independent methods to perform next-to-leading order (NLO) QCD calculations (see, e.g., Refs.~\cite{Frixione:1995ms, csdip, Frixione:1997np, Catani:2002hc}). Similarly, soft/collinear factorization at ${\cal O}(\as^2)$ \cite{Campbell:1997hg, Catani:1998nv, Bern:1998sc, Kosower:1999rx, Bern:1999ry, Catani:1999ss, Catani:2000pi, Czakon:2011ve, Bierenbaum:2011gg, Czakon:2018iev, Catani:2011st, Sborlini:2013jba} is used to
set up and develop methods (see, e.g., the reviews in Refs.~\cite{Heinrich:2020ybq, Proceedings:2018jsb, Amoroso:2020lgh, TorresBobadilla:2020ekr}) 
at the next-to-next-to-leading order (NNLO).


The knowledge of 
soft and collinear factorization of scattering amplitudesat ${\cal O}(\as^3)$ can be exploited in the context of N$^3$LO calculations and of resummed calculations at N$^3$LL accuracy. The singular factors for the various collinear limits at  ${\cal O}(\as^3)$ were presented in Refs.~\cite{DelDuca:1999iql, Birthwright:2005ak, Birthwright:2005vi,
DelDuca:2019ggv, DelDuca:2020vst,
 Catani:2003vu, Sborlini:2014mpa, Sborlini:2014kla, Badger:2015cxa, Czakon:2022fqi,
Bern:2004cz, Badger:2004uk, Duhr:2014nda, Catani:2011st}. 
The study of 
soft factorization of scattering amplitudes at ${\cal O}(\as^3)$ involves two-loop, one-loop
and tree-level contributions for various soft-parton multiplicities. 
Single soft-gluon emission at two loop order was examined in Refs.~\cite{Badger:2004uk, Li:2013lsa, Duhr:2013msa, Dixon:2019lnw}.
Double soft-parton radiation at one loop level was considered in Refs.~\cite{Zhu:2020ftr,Catani:2021kcy}.
 Triple soft-gluon emission at the tree level was studied in Ref.~\cite{Catani:2019nqv}. 
This paper is devoted to study soft gluon-quark-antiquark ($gq{\bar q}$) radiation at the tree level, which has been independently considered very recently in 
Ref.~\cite{DelDuca:2022noh}. Comments on Ref.~\cite{DelDuca:2022noh} are presented throughout
the paper.



The outline of the paper is as follows.
In Sect.~\ref{s:sfsc} we first introduce our notation and recall the soft factorization formula for scattering amplitudes. 
Then 
we present the calculation of the tree-level current for soft $gq{\bar q}$
emission in a generic hard-scattering process.
The result for the current has an irreducible correlation component that includes contributions with both abelian and non-abelian characters. 
In Sect.~\ref{s:tlsc} we consider soft factorization of squared amplitudes
and we compute the squared current for soft $gq{\bar q}$ radiation.
The squared current leads to irreducible colour dipole and colour tripole interactions.
The colour tripole interactions are odd under charge conjugation and they produce charge asymmetry effects between the soft quark and antiquark. 
In Sect.~\ref{s:e23h} we consider the specific applications to 
processes with two and three hard partons and, in particular,
we discuss the structure of the corresponding charge asymmetry contributions.
In Sect.~\ref{s:qed} we generalize our QCD results for soft $gq{\bar q}$ emission
to the cases of QED and mixed QCD$\times$QED radiative corrections for soft 
photon-fermion-antifermion and gluon-fermion-antifermion emissions.
A brief summary of our results is presented in Sect.~\ref{s:conc}.
In Appendix~\ref{a:tthp} we list the action of colour tripole operators onto scattering amplitudes with two and three hard partons.

\section{Soft factorization and soft currents\label{s:sfsc}}

In this section we first introduce our notation, mostly
following the notation that is also used in Refs.~\cite{Catani:2019nqv, Catani:2021kcy}
(more details can be found therein). 
We also briefly recall 
the factorization properties of scattering amplitudes in the soft limit and the known 
tree-level results for the emission of one soft gluon and the emission of a soft 
quark--antiquark pair. 
Then
we present and discuss our results of the soft current for the emission of a
$gq{\bar q}$
system at the tree level.

\subsection{Soft factorization of scattering amplitudes\label{s:sfsa}}

We study the soft behaviour of a generic scattering amplitude $\M$ whose
external-leg particles are on shell and with physical spin polarizations. In our notation
all external particles of $\M$ are treated as `outgoing' particles
(although they can be initial-state and final-state physical particles), with corresponding outgoing momenta and quantum numbers (e.g., colour, spin and flavour).
The perturbative evaluation of $\M$ is performed by using dimensional regularization in
$d=4-2\epsilon$ space-time dimensions, and $\mu$ is the dimensional regularization scale. Specifically, we use conventional dimensional regularization (CDR), with $d-2$ spin polarization states for on shell gluons (and photons) 
and 2 polarization states for on shell massless quarks or antiquarks (and massless leptons). 

We consider the behaviour of 
$\M$ in the
kinematical configuration where one or more of the momenta of the external-leg
massless particles become soft. 
We denote the
soft momenta by $q_\ell^\mu$ ($\ell=1,\dots,N$, and $N$ is the total
number of soft particles), while the momenta of the hard particles in $\M$ are
denoted by $p_i^\mu$ (in general they are not massless and $p_i^2 \equiv m_i^2 \neq 0$)
In this kinematical configuration, $\M(\{q_\ell\},
\{p_i\})$ becomes singular. The dominant singular behaviour is given by the
following factorization formula in colour space 
\cite{Catani:1999ss, Bern:1999ry, Catani:2000pi}:
\begin{equation}\label{1gfact}
  \ket{\M(\{q_\ell\}, \{p_i\})} \simeq
\J(q_1,\cdots,q_N) \; \ket{\M (\{p_i\})} \;.
\end{equation}
Here $\M (\{p_i\})$ is
the scattering amplitude that is obtained from the original amplitude
$\M(\{q_\ell\}, \{p_i\})$ by simply removing the soft external legs.  The factor
$\J$ is the soft current for multi-particle radiation from the scattering
amplitude.

At the formal level the soft behaviour of $\M(\{q_\ell\}, \{p_i\})$ is specified
by performing an overall rescaling of all soft momenta as $q_\ell \to \xi
q_l$ (the rescaling parameter $\xi$ is the same for each soft momentum $q_\ell$)
and by considering the limit $\xi\to 0$.  In this limit, the amplitude is
singular and it behaves as $(1/\xi)^N$ (modulo powers of $\ln \xi$ from loop
corrections). This dominant singular behaviour is embodied in the soft current
$\J$ on the right-hand side of Eq.~(\ref{1gfact}).  In this equation the symbol
$\simeq$ means that on the right-hand side we neglect contributions that
are less singular than $(1/\xi)^N$ in the limit $\xi\to 0$.

The soft current $\J(q_1,\cdots,q_N)$ in Eq.~(\ref{1gfact}) depends on the
momenta, colours and spins of both the soft and hard partons in the scattering
amplitude (although the hard-parton dependence is not explicitly denoted in the
argument of $\J$). However this dependence entirely follows from the
external-leg content of $\M$, and the soft current is completely independent of
the internal structure of the scattering amplitude. In particular, we remark
that the factorization in Eq.~(\ref{1gfact}) is valid \cite{Bern:1999ry,
  Catani:2000pi, Feige:2014wja} at arbitrary perturbative orders in the loop expansion of the
scattering amplitude. 
Therefore on both sides of Eq.~(\ref{1gfact}) the scattering amplitudes have the loop
expansion ${\ket \M}= {\ket {\M^{(0)}}}+{\ket {\M^{(1)}}}+\dots$, where $\M^{(0)}$
is the contribution to $\M$ at the lowest perturbative order,
$\M^{(1)}$ is the one-loop contribution, and so forth.
Correspondingly, we have $\J= \J^{(0)}+\J^{(1)}+\dots$,
where $\J^{(n)}$ is the contribution to $\J$ at the $n$-th loop accuracy.  In
 the following sections of this paper we limit ourselves to considering 
explicit expressions of 
only tree-level currents $\J^{(0)}$ and, for the sake of
simplicity, we simply denote them by $\J$ (removing the explicit superscript
$(0)$).

Considering the emission of soft QCD partons,
the all-loop 
current $\J$ in Eq.~(\ref{1gfact}) is an operator that acts
from the colour+spin space of $\M(\{p_i\})$ to the enlarged space of
$\M(\{q_\ell\}, \{p_i\})$.  In particular, soft radiation produces colour
correlations.  To take into account the colour structure we use the colour (+
spin) space formalism of Ref.~\cite{csdip}. The scattering amplitude
$\M_{s_1 s_2 \dots}^{c_1 c_2\dots}$
depends on the colour ($c_i$) and
spin 
($s_i$)
indices of its external-leg partons. This dependence is
embodied in a vector $\ket{\M}$ in colour+spin space through the definition
(notation)
\begin{equation}\label{Mstate}
  \M_{s_1 s_2 \dots}^{c_1 c_2 \dots} \equiv
  \big(\bra{c_1,c_2,\cdots}\otimes\bra{s_1,s_2,\cdots}\big) \;
  \ket{\M} \;\;,
\end{equation}
where 
$\{ \,\ket{c_1,c_2,\cdots}\otimes\ket{s_1,s_2,\cdots} \} = \{ \,
\ket{c_1,s_1;c_2,s_2,\cdots} \}$ 
is an orthonormal basis of abstract
vectors in colour+spin space.

In colour space the colour correlations produced by soft-gluon emission are
represented by associating a colour charge operator $\T_i$ to the emission of a
gluon from each parton $i$. If the emitted gluon has colour index $a$
($a=1,\dots,N_c^2-1$, for $SU(N_c)$ QCD with $N_c$ colours) in the adjoint
representation, the colour charge operator is $\T_i \equiv \bra{a} \,T_i^a$ and
its action onto the colour space is defined by
\begin{equation}\label{defT}
  \bra{a,c_1,\cdots,c_i,\cdots,c_m}\,\T_i\,\ket{b_1,\cdots,b_i,\cdots,b_m} \equiv
  \delta_{c_1 b_1} \cdots (T^a)_{c_i b_i} \cdots \delta_{c_m b_m} \;,
\end{equation}
where the explicit form of the colour matrices $T^a_{c_i b_i}$ depends on the
colour representation of the parton $i$:
\begin{align}
 (T^a)_{b c} &= \ui f^{bac} && \text{(adjoint representation) 
if $i$ is a gluon,} &&\nonumber \\
  (T^a)_{\alpha\beta} &=  t^a_{\alpha\beta} &&
  \text{(fundamental representation with $\alpha,\beta=1,\dots,N_c$)
    if $i$ is a quark,}  &&\nonumber \\
  (T^a)_{\alpha\beta} &= -t^a_{\beta\alpha}   &&
  \text{if $i$ is an antiquark.} &&\nonumber &&\label{Tcs}
\end{align}
We normalize the colour matrices such as $[T_i^a , T_j^b] = i f^{abc} T_i^c \delta_{ij}$
and ${\rm Tr} (t^a t^b) = T_R \,\delta_{ab}$ with $T_R=1/2$.
We also use the notation
$\sum_a T_i^a T_k^a \equiv \T_i \cdot \T_k$ and $\T_i^2 = C_i$, where $C_i$ is the
quadratic Casimir coefficient of the colour representation, with the
normalization $C_i=C_A=N_c$ if $i$ is a gluon and $C_i=C_F=(N_c^2-1)/(2N_c)$ if
$i$ is a quark or antiquark.

Note that each `amplitude vector' $\ket{\M}$ is an overall colour-singlet state.
Therefore, colour conservation is simply expressed by the relation
\begin{equation}
\label{colcons}
\sum_i \;\T_i \;\ket{\M} = 0 \;\;,
\end{equation}
where the sum extends over all the external-leg partons $i$ of the amplitude $\M$. 
For subsequent use, we also introduce the shorthand notation
\begin{equation}
\label{csnotation}
\sum_i \;\T_i \;\eqcs \;0 \;\;,
\end{equation}
where the subscript CS in the symbol $\eqcs$ means that the equality between the
terms in the left-hand and right-hand sides of the equation is valid if these
(colour operator) terms act (either on the left or on the right) onto
colour-singlet states.

\subsection{Tree-level currents\label{s:tlc}}

The tree-level current $\bj(q)$ for the emission of a single soft gluon of momentum
$q^\nu$ is well known~\cite{Bassetto:1984ik}:
\begin{equation}\label{J1}
\bj(q) = \g \,\mu^\ep 
\;\sum_i  
\;\bt_i \;
\frac{p_i \cdot \pol(q)}{p_i \cdot q} 
\equiv \bj_\nu(q) \pol^\nu(q)\;\;,
\end{equation}
where $\g$ is the QCD coupling ($\as=\g^2/(4\pi)$). The notation $\sum_i$
means that the sum extends over all hard partons
(with momenta $p_i$) in $\cm$, $\bt_i$ is the colour charge of the hard parton $i$, and
$\varepsilon^\nu(q)$ is the spin polarization 
vector of the soft gluon.

The current for emission of soft gluons is conserved by acting on colour-singlet states
(see Ref.~\cite{Catani:2019nqv} for a general discussion on soft-current conservation).
From Eq.~(\ref{J1}) we have
\begin{equation}\label{consJ1}
  q^\nu \,\bj_\nu(q) = \sum_i \bt_i \;\;,
\end{equation}
and, therefore, by using colour conservation in Eq.~\eqref{colcons},
the current conservation relation $q^\nu \,\bj_\nu(q) \eqcs 0$
is directly fulfilled.

The tree-level currents for emission of two and three soft gluons were computed in 
Refs.~\cite{Catani:1999ss} and \cite{Catani:2019nqv}, respectively.

The emission of a soft quark-antiquark ($q{\bar q}$) pair by tree-level QCD interactions was studied
in Ref.~\cite{Catani:1999ss}. Using our notation, the QCD current for radiation of a soft quark and antiquark at the tree level is \cite{Catani:2021kcy}
\begin{equation}\label{Jqa}
\bj(q_1,q_2) 
=
- \left( \g\,\mu^\ep\right)^2 \,
\sum_i 
\,\btq^c \;T^c_i
\;\frac{p_i \cdot j(1,2)}{p_i \cdot q_{12}} \;\;,
\end{equation}
where we have introduced the fermionic current $j^\nu(1,2)$,
\begin{equation}\label{fercur}
j^\nu(1,2) \equiv \frac{\ubar(q_1)\, \gamma^\nu \,v(q_2)}{q_{12}^2} \;\;,
\quad \quad \;\;\; \quad q_{12} = q_1+q_2 \;\;.
\end{equation}
The soft quark and antiquark have momenta $q_1^\nu$ and $q_2^\nu$, respectively,
and $u(q)$ and $v(q)$ are the customary Dirac spinors. The spin indices ($s_1$ and
$s_2$) and the colour indices ($\alpha_1$ and $\alpha_2$) of the 
quark and antiquark are embodied in the colour+spin space notation of 
Eq.~(\ref{Jqa}). 
Considering the projection 
$(\bra{\alpha_1,\alpha_2} \otimes \bra{s_1,s_2} \,) \,\bj(q_1,q_2)
\equiv J^{\alpha_1,\alpha_2}_{s_1,s_2}(q_1,q_2)$
of the current onto its colour and spin indices, we have
$(\bra{\alpha_1,\alpha_2} \otimes \bra{s_1,s_2} \,) \,\btq^c \;
\ubar(q_1)\, \gamma^\nu \,v(q_2) = t^c_{\alpha_1\alpha_2} \,
\ubar_{(s_1)}(q_1)\, \gamma^\nu \,v_{(s_2)}(q_2)$.

\subsubsection{The tree-level current for soft $gq{\bar q}$ emission\label{s:gqqc}}

\begin{figure}
\begin{center}
\fcolorbox{white}{white}{
  \begin{picture}(300,200) (155,-40)
  \scalebox{.55}{
     \SetColor{Black}
    \Text(194,140)[]{\Large A)}
    \Text(370,260)[]{\Large $q_1$}
    \Text(370,222)[]{\Large $q_2$}
    \Text(370,158)[]{\Large $q_3$}
    \Text(240,240)[]{\Large $p_i$}
    \Text(240,200)[]{\Large $p_j$}
    \SetWidth{1.0}
    \Gluon(255,256)(351,256){4.5}{8}
    \Gluon(255,184)(303,188){4.5}{4}
    \SetWidth{1.9}
    \Line(207,244)(303,268)
    \Line(207,196)(303,172)
    \SetWidth{1.0}
    \GOval(183,220)(36,36)(0){0.882}
    \Gluon(255,256)(351,256){4.5}{8}
    \Line[arrow,arrowpos=0.5,arrowlength=5,arrowwidth=2,arrowinset=0.2](303,189)(352,220)
    \Line[arrow,arrowpos=0.5,arrowlength=5,arrowwidth=2,arrowinset=0.2,flip](303,188)(351,160)
    %
    \Text(488,140)[]{\Large B)}
    \Text(604,300)[]{\Large $q_1$}
    \Text(690,300)[]{\Large $q_2$}
    \Text(690,266)[]{\Large $q_3$}
    \Text(580,190)[]{\Large $p_i$}
    \SetWidth{1.0}
    \GOval(481,222)(36,36)(0){0.882}
    \SetWidth{1.9}
    \Line(517,222)(649,222)
    \SetWidth{1.0}
    \Gluon(541,222)(601,282){4.5}{7}
    \Gluon(601,222)(637,257){4.5}{4}
    \Line[arrow,arrowpos=0.5,arrowlength=5,arrowwidth=2,arrowinset=0.2](637,258)(673,282)
    \Line[arrow,arrowpos=0.5,arrowlength=5,arrowwidth=2,arrowinset=0.2,flip](637,258)(673,258)
    %
    \Text(799,140)[]{\Large C)}
    \Text(977,300)[]{\Large $q_1$}
    \Text(935,300)[]{\Large $q_2$}
    \Text(935,266)[]{\Large $q_3$}
    \Text(910,190)[]{\Large $p_i$}
    \SetWidth{1.0}
    \Gluon(850,223)(886,258){4.5}{4}
    \Line[arrow,arrowpos=0.5,arrowlength=5,arrowwidth=2,arrowinset=0.2](886,259)(922,283)
    \Line[arrow,arrowpos=0.5,arrowlength=5,arrowwidth=2,arrowinset=0.2,flip](886,259)(922,259)
    \SetWidth{1.0}
    \GOval(792,222)(36,36)(0){0.882}
    \SetWidth{1.9}
    \Line(828,222)(960,222)
    \SetWidth{1.0}
    \Gluon(913,223)(973,283){4.5}{7}
    %
    \Text(194,-50)[]{\Large D)}
    \Text(370,120)[]{\Large $q_2$}
    \Text(370,80)[]{\Large $q_1$}
    \Text(370,38)[]{\Large $q_3$}
    \Text(282,-7)[]{\Large $p_i$}
    \SetWidth{1.0}
    \GOval(184,31)(36,36)(0){0.882}
    \SetWidth{1.9}
    \Line(220,31)(353,31)
    \SetWidth{1.0}
    \Gluon(232,31)(280,91){4.5}{6}
    \Line[arrow,arrowpos=0.5,arrowlength=5,arrowwidth=2,arrowinset=0.2,flip](280,91)(352,43)
    \Line[arrow,arrowpos=0.5,arrowlength=5,arrowwidth=2,arrowinset=0.2](280,91)(352,127)
    \Gluon(304,103)(352,75){4.5}{4}
    %
    \Text(488,-50)[]{\Large E)}
    \Text(580,-7)[]{\Large $p_i$}
    \Text(665,95)[]{\Large $q_1$}
    \Text(665,120)[]{\Large $q_2$}
    \Text(665,38)[]{\Large $q_3$}
    \SetWidth{1.0}
    \GOval(481,30)(36,36)(0){0.882}
    \SetWidth{1.9}
    \Line(517,30)(650,30)
    \SetWidth{1.0}
    \Gluon(529,30)(577,90){4.5}{6}
    \Line[arrow,arrowpos=0.5,arrowlength=5,arrowwidth=2,arrowinset=0.2,flip](577,90)(649,42)
    \Line[arrow,arrowpos=0.5,arrowlength=5,arrowwidth=2,arrowinset=0.2](577,90)(649,126)
    \Gluon(601,75)(649,96){4.5}{4}
    %
    \Text(799,-50)[]{\Large F)}
    \Text(977,50)[]{\Large $q_1$}
    \Text(977,120)[]{\Large$q_2$}
    \Text(977,80)[]{\Large $q_3$}
    \Text(910,-7)[]{\Large $p_i$}
    \SetWidth{1.0}
    \GOval(793,31)(36,36)(0){0.882}
    \SetWidth{1.9}
    \Line(829,31)(961,31)
    \SetWidth{1.0}
    \Gluon(853,31)(877,67){4.5}{3}
    \Gluon(877,67)(913,103){4.5}{4}
    \Gluon(877,67)(961,67){4.5}{7}
    \Line[arrow,arrowpos=0.5,arrowlength=5,arrowwidth=2,arrowinset=0.2](913,103)(961,126)
    \Line[arrow,arrowpos=0.5,arrowlength=5,arrowwidth=2,arrowinset=0.2,flip](913,103)(961,79)
    }
  \end{picture}
}
\caption{
\label{fig:tree}
{\em Feynman diagrams that contribute to the current $\bj(q_1,q_2,q_3)$ for soft 
$gq{\bar q}$ emission. The external-leg hard
partons with momenta $p_i$ and $p_j$ are coupled to gluons by using the
eikonal approximation. 
The scattering amplitude $\M(\{p_i\})$ is denoted by the 
grey circle.
}}
\end{center}
\end{figure}
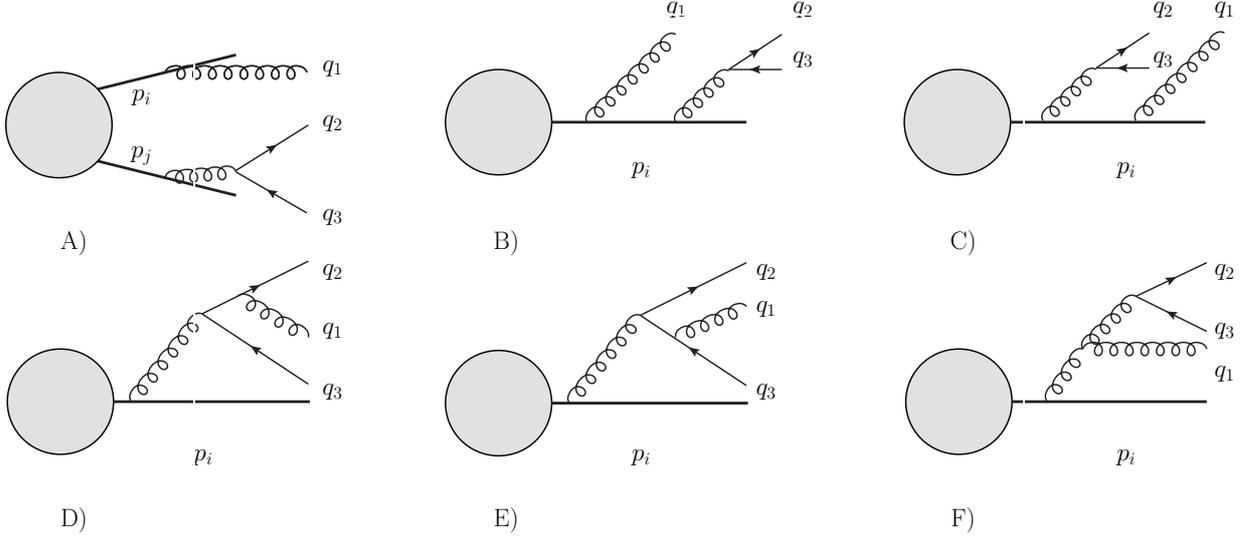

The tree-level current for soft $gq{\bar q}$ emission is denoted by $\bj(q_1,q_2,q_3)$.
The soft gluon has momentum $q_1$, while $q_2$ and $q_3$ are the momenta of the soft quark and antiquark, respectively. 

We compute $\bj(q_1,q_2,q_3)$ by using the method of Ref.~\cite{Catani:1999ss}, namely,
we consider eikonal emission of the three soft partons from the external hard partons of the generic scattering amplitude $\M$. The relevant Feynman diagrams are shown in 
Fig.~\ref{fig:tree}.
The external-leg hard partons with momenta $p_i$ and $p_j$ are coupled to gluons by using the eikonal approximation for both vertices and propagators.
The remaining contributions to the Feynman diagrams in 
Fig.~\ref{fig:tree}
are treated without any approximations for vertices and propagators. 
We note that the propagators of the off shell (internal-line) gluons are gauge dependent.
We have computed the current by using both axial and covariant gauges for the polarization tensor of the internal-line gluons, and we have checked that the final result for 
$\bj(q_1,q_2,q_3)$ is explicitly gauge independent. More precisely, the total contribution
of the gauge dependent terms vanishes by using the colour conservation relation in 
Eq.~(\ref{colcons}).

We present our result for $\bj(q_1,q_2,q_3)$ in the following form:
\begin{equation}\label{J123}
  \bj(q_1,q_2,q_3) 
= 
\sy{\bj(q_1)}{\bj(q_2,q_3)} + \bs\Gamma(q_1,q_2,q_3) \;\;,
\end{equation}
where $\bj(q_1)$ and $\bj(q_2,q_3)$ are the currents in Eqs.~(\ref{J1}) and (\ref{Jqa}),
and we have introduced the symbol $(\dots)_{sym}$ to denote symmetrized products.
The symmetrized product of two colour space operators $A$ and $B$ is defined as
\begin{equation}\label{symprod}
  \sy{A}{B} \equiv \frac12(AB+BA) \;.
\end{equation}
The right-hand side of Eq.~(\ref{J123}) has the structure of an expansion in irreducible
correlations, which is analogous to the structure of the two-gluon and three-gluon
soft currents in Refs.~\cite{Catani:1999ss} and \cite{Catani:2019nqv}, respectively.
The first term in the right-hand side of Eq.~(\ref{J123}) represents the `independent'
(though colour-correlated) emission of the soft gluon and the soft $q{\bar q}$ pair
from the hard partons. The term $\bs\Gamma(q_1,q_2,q_3)$ is definitely an irreducible correlation contribution to soft $gq{\bar q}$ emission.

To present our result for the irreducible contribution
we consider the projection $\bj^{a_1} \equiv \bra{a_1} \bj$ of the current onto the colour index $a_1$ of the soft gluon. The corresponding projection 
$\bs\Gamma^{a_1} \equiv \bra{a_1} \bs\Gamma$ of the irreducible correlation has the following explicit expression:
\begin{equation}
  \bs\Gamma^{a_1}(q_1,q_2,q_3) = \left( \g\,\mu^\ep\right)^3 \,
  \sum_i T_i^c \;\bs\gamma_i^{a_1 c}(q_1,q_2,q_3) \;\;, \label{bfgamma} 
\end{equation}
where
\begin{align}
  \bs\gamma_i^{a_1 c}(q_1,q_2,q_3) & \equiv
   \frac12 \{\btq^{a_1},\btq^c\} \gab_i(q_1,q_2,q_3)
  +\frac12 [ \btq^{a_1},\btq^c ] \gna_i(q_1,q_2,q_3) \;\;, \label{gamma} \\
  \gab_i(q_1,q_2,q_3) &= \frac{\pol_\mu(q_1)}{q_{123}^2\;p_i\cdot q_{123}} \ub(q_2)\left(
   \sla{p}_i\frac{\sla{q}_{13}}{q_{13}^2}\gamma^\mu
   -\gamma^\mu\frac{\sla{q}_{12}}{q_{12}^2}\sla{p}_i\right)v(q_3) \;\;, \label{psi}\\
  \gna_i(q_1,q_2,q_3) &= \frac{\pol_\mu(q_1)}{\,p_i\cdot q_{123}} \;
  \ub(q_2)\left\{ \frac{p_i^\mu}{q_{23}^2}\left(\frac1{p_i\cdot q_1}
  -\frac1{p_i\cdot q_{23}}\right)\sla{p}_i\right. \label{chi} \\
  &\left. \!\!\! \!\!\!\!\!\!
  +\frac1{q_{123}^2}\left[\frac1{q_{23}^2}
    \left(2p_i\cdot(q_{23}-q_1)\gamma^\mu-4q_{23}^\mu\sla{p}_i +4p_i^\mu\sla{q}_1\right)
    -\gamma^\mu\frac{\sla{q}_{12}}{q_{12}^2}\sla{p}_i
    -\sla{p}_i\frac{\sla{q}_{13}}{q_{13}^2}\gamma^\mu
    \right]
  \right\}v(q_3) \;, \nn
\end{align}
and we have defined $q_{ij}=q_i + q_j$ and $q_{123}=q_1 + q_2 + q_3$.
The expression of $\bs\gamma_i^{a_1 c}$ in Eq.~(\ref{gamma}) involves products 
(an anticommutator and a commutator)
of two matrices $\btq^b$ in the fundamental representation. Considering the projection
$J^{a_1}_{\alpha_2 \alpha_3} \equiv \bra{a_1, \alpha_2,\alpha_3} \,\bj$ of the current
onto the colour indices $\alpha_2$ and $\alpha_3$ of the soft quark and antiquark,
the action of the matrices $\btq^b$ in Eq.~(\ref{gamma}) is
$\bra{\alpha_2,\alpha_3} \, \btq^{a} \btq^c = (t^a t^c)_{\alpha_2 \alpha_3}$.

The `independent' emission contribution $\sy{\bj(q_1)}{\bj(q_2,q_3)}$ in Eq.~(\ref{J123})
embodies products of the type $T_i^a T_k^b$ of colour charges of two hard partons.
In contrast, the irreducible component $\bs\Gamma$ in Eq.~(\ref{bfgamma})
has a linear dependence on the colour charges $T_i^c$ of the hard partons.
This feature of $\bs\Gamma$ is analogous to the linear dependence on $T_i^c$
of the irreducible component of the currents
for double \cite{Catani:1999ss} and triple \cite{Catani:2019nqv} soft-gluon emission.

We note that the irreducible component $\bs\Gamma$ in Eqs.~(\ref{bfgamma}) and 
(\ref{gamma}) embodies a contribution of abelian type, which is proportional
to the kinematical function $\gab_i$, in addition to a purely non-abelian contribution,
which is proportional to $\gna_i$.
In contrast, in the case of double and triple soft-gluon emission 
\cite{Catani:1999ss, Catani:2019nqv} the irreducible correlations are maximally non-abelian.
The presence of an abelian-type contribution in  Eqs.~(\ref{bfgamma}) and 
(\ref{gamma}) implies corresponding irreducible correlations in the current for soft
photon-lepton-antilepton emission in QED (see Sect.~\ref{s:qed}).

Writing $\bj(q_1,q_2,q_3) = \pol^\nu(q_1) \,\bj_\nu(q_1,q_2,q_3)$
we make explicit the dependence of the current on the Lorentz index $\nu$ of the soft gluon. It is straightforward to check that the result in Eqs.~(\ref{J123}) and 
(\ref{bfgamma})--(\ref{chi})
fulfils current conservation, namely,
\begin{equation}\label{currCons}
  q_1^\nu \,\bj_\nu(q_1,q_2,q_3) \eqcs 0 \;.
\end{equation}

The result of $\bj(q_1,q_2,q_3)$ in Eqs.~(\ref{J123})--(\ref{chi}) fully agrees with the corresponding result of the soft $gq{\bar q}$ current that was first computed in 
Ref.~\cite{DelDuca:2022noh}. Our results and those in Ref.~\cite{DelDuca:2022noh} formally
differ only in the presentation of the expression of $\bs\gamma_i^{a_1 c}$ in 
Eq.~(\ref{gamma}). We use the two independent colour structures $\{\btq^{a_1},\btq^c\}$
and $[\btq^{a_1},\btq^c]$, while Ref.~\cite{DelDuca:2022noh} uses the three colour structures
$\btq^{a_1} \btq^c$, $\btq^{c} \btq^{a_1}$ and $[\btq^{a_1},\btq^c]= i f^{a_1 c b} \btq^b$,
which are linearly dependent.

\section{Tree-level squared currents\label{s:tlsc}}

Using the colour+spin space notation of Sect.~\ref{s:sfsa}, 
the squared amplitude $|\M|^2$ (summed
over the colours and spins of its external legs) is written as follows
\beq
\label{squared}
|\M|^2 = \langle{\M} \ket{\,\M} \;\;.
\eeq
Accordingly, the square of the soft factorization formula (\ref{1gfact})
gives
\beq
\label{softsquared}
| \M(\{q_\ell\}, \{p_i\}) |^2 \simeq 
\bra{\M (\{p_i\})} \;| \J(q_1,\cdots,q_N) |^2 \;\ket{\M (\{p_i\})} \;,
\eeq
where
\beq
\label{spin}
| \J(q_1,\cdots,q_N) |^2 = 
\sum_{\{c_i\}, \{s_i\}}
\left[ J^{c_1 \dots c_N}_{s_1 \dots s_N}(q_1,\cdots,q_N) 
\right]^\dagger J^{c_1 \dots c_N}_{s_1 \dots s_N}(q_1,\cdots,q_N) \;\;,
\eeq
and, analogously to Eq.~(\ref{1gfact}), the symbol $\simeq$ means that we neglect contributions that are subdominant in the soft limit.
The squared current $| \J(q_1,\cdots,q_N) |^2 $, which is summed over the colours 
$c_1 \dots c_N$ and spins $s_1 \dots s_N$ of the soft partons, is still a colour operator that depends on the colour charges of the hard partons in $\M (\{p_i\})$. These colour charges produce colour correlations and,
therefore, the right-hand side of Eq.~(\ref{softsquared}) is not proportional to
$|\M(\{p_i\})|^2$ in the case of a generic scattering amplitude\footnote{Colour correlations can be simplified in the case of scattering amplitudes with two and three hard partons (see Sect.~\ref{s:e2h} and \ref{s:e3h}).}. 

The computation of the squared current in Eq.~(\ref{spin}) involves the sum over the physical spin polarization vectors of the soft gluons. These polarization vectors are gauge dependent, but the action of $| \J |^2 $ onto colour singlet states is fully gauge invariant,
since the gauge dependent contributions cancel as a consequence of the conservation of the soft current (see, e.g., Ref.~\cite{Catani:2019nqv}).


We recall the known results of the squared currents for emission of one soft gluon
and of a soft $q{\bar q}$  pair.
The square of the soft current $\bj(q_1)$ in Eq.~(\ref{J1}) for single gluon
emission is
\begin{equation}\label{J1sq}
  |\bj(q_1)|^2 \equiv \bj(q_1)^\dagger \bj(q_1)
  \eqcs  -\left( \g\,\mu^\ep\right)^2
  \sum_{i,k} \T_i\cdot \T_k \; \cS_{ik}(q_1) 
\;\;,
\end{equation}
where
\begin{equation}
\label{S1}
\cS_{ik}(q) = \frac{p_i\cdot p_k}{p_i\cdot q\; p_k\cdot q} \;\;.
\end{equation}
The square of the soft current $\bj(q_2,q_3)$ in Eq.~(\ref{Jqa}) for 
$q{\bar q}$
emission is \cite{Catani:1999ss}
\begin{equation}\label{Jqasq}
  |\bj(q_2,q_3)|^2 \equiv  \bj(q_2,q_3)^\dagger\; \bj(q_2,q_3)
  \eqcs \left( \g\,\mu^\ep\right)^4 T_R \sum_{i,k} \T_i\cdot \T_k \; {\cal I}_{ik}(q_2,q_3)
\;\;,
\end{equation}
where
\begin{equation}
 {\cal I}_{ik}(q_2,q_3) = \frac{p_i\cdot q_2\; p_k\cdot q_3+p_i\cdot q_3\;
   p_k\cdot q_2-p_i\cdot p_k \;q_2\cdot q_3}%
 {(q_2\cdot q_3)^2\;p_i\cdot q_{23}\; p_k\cdot q_{23}}
 \;. \label{I23}
\end{equation}

The colour charge dependence of both squared currents in Eqs.~(\ref{J1sq}) and (\ref{Jqasq}) 
is given in terms of dipole operators $\T_i\cdot \T_k = \sum_a T_i^a  T_k^a$.
The insertion of the dipole operators in the factorization formula (\ref{softsquared})
produces colour correlations between two hard partons ($i$ and $k$) in $\M(\{p_i\})$.

Using colour charge conservation (see Eq.~\eqref{colcons}), the single-gluon and $q\bar q$ squared
currents in Eqs.~(\ref{J1sq}) and (\ref{Jqasq}) 
can be rewritten as follows
\begin{align}
  |\bj(q_1)|^2 &\eqcs - \left( \g\,\mu^\ep\right)^2
  \, \frac{1}{2} \sum_{i \neq k}
  \T_i\cdot \T_k \, w_{ik}(q_1) \;\;,\label{j1q} \\
  w_{ik}(q_1) &= \cS_{ik}(q_1)+\cS_{ki}(q_1)-\cS_{ii}(q_1)-\cS_{kk}(q_1) \;\;,
  \label{w1} \\
  |\bj(q_2,q_3)|^2 &\eqcs - \left( \g\,\mu^\ep\right)^4 T_R
  \, \frac{1}{2} \sum_{i \neq k} 
  \T_i\cdot \T_k \, w_{ik}(q_2,q_3) \;\;,\label{j23q} \\
  w_{ik}(q_2,q_3) &= {\cal I}_{ii}(q_2,q_3)+{\cal I}_{kk}(q_2,q_3)
  -{\cal I}_{ik}(q_2,q_3)-{\cal I}_{ki}(q_2,q_3) \;\;. \label{w23}
\end{align}
As noticed in Refs.~\cite{Catani:2019nqv, Catani:2021kcy}, the expressions in 
Eqs.~(\ref{j1q}) and (\ref{j23q}) have a more straightforward physical interpretation,
since the kinematical functions $w_{ik}(q_1)$ in Eq.~(\ref{w1}) and
$w_{ik}(q_2,q_3)$ in Eq.~(\ref{w23}) are directly related to the intensity of soft radiation
from two distinct hard partons, $i$ and $k$, in a colour singlet configuration
(see Sect.~\ref{s:e2h}).

\subsection{The squared current for soft $gq{\bar q}$ radiation}\label{s:gqqsc}

The soft-$q\bar q$ kinematical functions ${\cal I}_{ik}(q_2,q_3)$ and $w_{ik}(q_2,q_3)$
in Eqs.~(\ref{I23}) and (\ref{w23}) are {\it symmetric} with respect to the exchange
$q_2 \leftrightarrow q_3$ of the quark and antiquark momenta. The evaluation of the one-loop
QCD corrections to the soft-$q\bar q$ squared current \cite{Catani:2021kcy} produces also
kinematical correlations with an {\it antisymmetric} dependence with respect to
$q_2 \leftrightarrow q_3$. As discussed in Ref.~\cite{Catani:2021kcy}, such
antisymmetric dependence is related to charge asymmetry effects that are distinct features
of the radiation of quarks and antiquarks. An antisymmetric dependence with respect to
$q_2 \leftrightarrow q_3$ and ensuing charge asymmetry effects occur also in the tree-level squared current $|\bj(q_1,q_2,q_3)|^2$ for soft $gq\bar q$ radiation, as we discuss
in the following. Similar to  Ref.~\cite{Catani:2021kcy}, the charge asymmetry effects
for soft $gq\bar q$ radiation involve colour correlations that depend on the fully-symmetric
colour tensor\footnote{If $N_c=2$ the tensor $d^{abc}$ vanishes and there are no
charge asymmetry effects.} $d^{abc}$,
\begin{equation}
\label{dtens}
d^{abc} = \frac{1}{T_R} \,{\rm Tr}\left( \left\{t^a, t^b \right\} t^c \right) \;\;,
\end{equation}
which is odd under charge conjugation.

We compute the square of the tree-level $gq\bar{q}$ current by using the structure in
Eq.~\eqref{J123} and we obtain the following three contributions:
{\it (I)} the square of the `uncorrelated' current
$\sy{\bj(q_1)}{\bj(q_2,q_3)}$, {\it (II)} the product of the `uncorrelated' current
and the irreducible current $\bs\Gamma(q_1,q_2,q_3)$, {\it (III)} the square of the irreducible current.

The square of the `uncorrelated' current
$\sy{\bj(q_1)}{\bj(q_2,q_3)}$
can be written in terms of the symmetric product
of the squared currents $|\bj(q_1)|^2$ and $|\bj(q_2,q_3)|^2$, and an additional
irreducible term $W^{(I)}$, which involves colour dipole correlations. We obtain
\begin{align}\label{Jsq1}
  &|\sy{\bj(q_1)}{\bj(q_2,q_3)}|^2
  \eqcs \sy{|\bj(q_1)|^2}{|\bj(q_2,q_3)|^2} + W^{(I)}(q_1,q_2,q_3) \;\;, \\
  &W^{(I)}(q_1,q_2,q_3) = -\left( \g\,\mu^\ep\right)^6
  T_R C_A \sum_{i,k} \T_i\cdot \T_k \; \cS^{(I)}_{ik}(q_1,q_2,q_3) \;\;,
  \label{WI}
\end{align}
where the momentum function $\cS^{(I)}_{ik}(q_1,q_2,q_3)$ is\footnote{In Eq.~(\ref{SI})
and in the following equations the scalar products $v \cdot u$ between to generic
momenta $u^\mu$ and $v^\mu$ are simply denoted by $vu$.}
\begin{align}
  &\cS^{(I)}_{ik}(q_1,q_2,q_3) =
  \frac{\pq{i}{2}}{4 \per (\qq{2}{3})^2 \per \pq{i}{1} \per \pq{i}{23}} \per
  \left[\frac{\pp{i}{k}}{\pq{k}{1}} \per \left(\frac{3 \per
        \pq{k}{3}}{\pq{k}{23}}-\frac{2 \per
        \pq{i}{3}}{\pq{i}{23}}\right)-\frac{2 \per m_i^2 \per
      \pq{k}{3}}{\pq{i}{1} \per \pq{k}{23}}\right] \nonumber \\
  &+\frac{\pp{i}{k}}{8 \per \qq{2}{3} \per \pq{i}{1} \per \pq{i}{23}} \per
  \left[\frac{-3 \per \pp{i}{k}}{\pq{k}{1} \per \pq{k}{23}}+2 \per m_i^2 \per
    \left(\frac{1}{\pq{i}{1} \per \pq{k}{23}}+\frac{1}{\pq{k}{1} \per
        \pq{i}{23}}\right)\right] +(2\lra 3) \;\;.
  \label{SI}
\end{align}

The product of the `uncorrelated' current and the irreducible current leads to colour correlations that involve dipoles and tripoles. We have
\begin{align}
  &\sum\sy{\bj(q_1)}{\bj(q_2,q_3)}^\dagger\,\bs\Gamma(q_1,q_2,q_3) + \text{h.c.}
  = W^{(II)}(q_1,q_2,q_3)
  \nonumber \\
  &\eqcs  -\left( \g\,\mu^\ep\right)^6 \, T_R \left[
    C_A \sum_{i,k}\T_i\cdot \T_k \; \cS^{(II)}_{ik}(q_1,q_2,q_3)
    + \sum_{i,k,m} \tri{ikm} \;\cS_{ikm}(q_1,q_2,q_3)\right] \;, \label{W2}
\end{align}
where `h.c' denotes the hermitian-conjugate contribution, and
we have defined the (hermitian) $d$-type colour tripole  $\tri{ikm}$ as
\begin{equation}\label{tripoli}
  \tri{ikm} \equiv \sum_{a,b,c} d^{abc} T_i^a T_k^b T_m^c \;.
\end{equation}
The complete symmetry of $d^{abc}$ causes $\tri{ikm}$ to be completely symmetric
in the permutations of its indices $i, k, m$.

The kinematical coefficient $\cS^{(II)}_{ik}$ of the dipole contribution to Eq.~(\ref{W2})
reads
\begin{align}
 &\cS^{(II)}_{ik}(q_1,q_2,q_3) = 
   \frac{1}{2\qq{2}{3} \per \pq{i}{123}} \per \left\{\frac{2}{q_{123}^2 \per
     \qq{2}{3}} \per \left(\frac{\pq{k}{2} \per (2 \per m_i^2 \per
       \qq{1}{3}-\pq{i}{123} \per \pq{i}{3})}{\pq{i}{1} \per
       \pq{k}{23}} \right.\right. \nonumber\\
 & \qquad\qquad \left. -\frac{\pq{i}{2} \per (2 \per \pp{i}{k} \per
       \qq{1}{3}-(\pq{i}{1}-\pq{i}{2}+3 \per \pq{i}{3}) \per
       \pq{k}{3})}{\pq{k}{1} \per \pq{i}{23}}\right)\nonumber\\
&+\frac{1}{q_{123}^2 \per \qq{1}{2}} \per \left(\frac{2 \per \pq{i}{12} \per
    (\pp{i}{k} \per \qq{2}{3}-\pq{i}{2} \per \pq{k}{3}-\pq{i}{3} \per
    \pq{k}{2})+m_i^2 \per (\pq{k}{2} \per \qq{1}{3}-\pq{k}{1} \per
    \qq{2}{3})}{\pq{i}{1} \per \pq{k}{23}}\right.\nonumber\\
&\left.-\frac{m_i^2 \per ((\pq{k}{1}+2 \per \pq{k}{2}) \per \qq{2}{3}+\pq{k}{2}
    \per \qq{1}{3})-2 \per (\pq{i}{2} \per \pq{k}{1}+(\pq{i}{1}+2 \per
    \pq{i}{2}) \per \pq{k}{2}) \per \pq{i}{3}}{\pq{k}{1} \per
    \pq{i}{23}}\right)\nonumber\\
&+\frac{1}{q_{123}^2} \per \left(\frac{\pp{i}{k} \per \pq{i}{123}+m_i^2 \per
    (\pq{k}{2}-2 \per \pq{k}{1})}{\pq{i}{1} \per \pq{k}{23}}-\frac{3 \per m_i^2
    \per \pq{k}{2}-\pp{i}{k} \per \pq{i}{1}}{\pq{k}{1} \per
    \pq{i}{23}}\right)\nonumber\\
&+\frac{1}{\qq{2}{3}} \per \left(\pq{i}{2} \per
  \left(\frac{1}{\pq{i}{1}}-\frac{1}{\pq{i}{23}}\right) \per \left(\frac{m_i^2
      \per \pq{k}{3}}{\pq{i}{1} \per \pq{k}{23}}-\frac{\pp{i}{k} \per
      \pq{i}{3}}{\pq{k}{1} \per \pq{i}{23}}\right)\right)\nonumber\\
&\left.+\frac{1}{2} \per m_i^2 \per \pp{i}{k} \per
  \left(\frac{1}{\pq{i}{23}}-\frac{1}{\pq{i}{1}}\right) \per
  \left(\frac{1}{\pq{i}{1} \per \pq{k}{23}}-\frac{1}{\pq{k}{1} \per
      \pq{i}{23}}\right)\right\} + (2\lra 3) \;\;.
\label{SII}
\end{align}
The kinematical coefficient $\cS_{ikm}$ of the tripole contribution to Eq.~(\ref{W2})
is
\begin{align}
 &\cS_{ikm}(q_1,q_2,q_3) =
\frac{1}{\pq{i}{1} \per \pq{k}{23} \per \pq{m}{123} \per q_{123}^2}
\per \left\{\frac{1}{\qq{1}{2} \per \qq{2}{3}} \per \Big[\pq{i}{1} \per (\pq{k}{2}
  \per \pq{m}{3}+\pq{k}{3} \per \pq{m}{2}) \right. \nonumber\\
 & +\pq{i}{2} \per \big(\pq{k}{1} \per
  \pq{m}{3}+\pq{k}{3} \per \pq{m}{1} +2 \per (\pq{k}{2} \per \pq{m}{3}+\pq{k}{3}
  \per \pq{m}{2}) \big) \nonumber\\
  &+\pq{i}{3} \per(-\pq{k}{1} \per \pq{m}{2}+\pq{k}{2} \per \pq{m}{1})+\qq{1}{3} \per (\pp{i}{k}
\per \pq{m}{2}-\pp{i}{m} \per \pq{k}{2}-\pp{k}{m} \per \pq{i}{2})\Big] \nonumber\\
&+\frac{1}{\qq{1}{2}} \per \big(\pp{i}{m} \per \pq{k}{1}-\pp{i}{k} \per
  \pq{m}{1}-\pp{k}{m} \per (\pq{i}{1}+2 \per \pq{i}{2})\big) \nonumber\\
 & \left.+\frac{1}{\qq{2}{3}}
  \per (\pp{k}{m} \per \pq{i}{3}-\pp{i}{k} \per \pq{m}{3}-\pp{i}{m} \per
  \pq{k}{3})\right\} - (2\lra 3)\;. \label{Sikm}
\end{align}
Note that, while all dipole kinematical coefficients are symmetric in the exchange 
$q_2 \leftrightarrow q_3$ of the quark
and antiquark momenta (see also $\cS_{ik}^{(III,\ab)}$ and $\cS_{ik}^{(III,\na)}$
in Eqs.~(\ref{sIIIab}) and (\ref{sIIIna})), the coefficient 
$\cS_{ikm}(q_1,q_2,q_3)$ of the tripole is
antisymmetric in such exchange.

The square of the irreducible current gives
\begin{align}
  &\bs\Gamma^\dagger(q_1,q_2,q_3) \, \bs\Gamma(q_1,q_2,q_3) = W^{(III)}(q_1,q_2,q_3) \nonumber \\
  &\eqcs -\left( \g\,\mu^\ep\right)^6\, T_R \sum_{i,k} \left\{
  \left(C_F-\frac{C_A}{4}\right)
  \;\cS^{(III,\ab)}_{ik}(q_1,q_2,q_3) + \frac{C_A}{4}\;\cS^{(III,\na)}_{ik}(q_1,q_2,q_3)
  \right\} \T_i\cdot \T_k \;. \label{Gammasq}
\end{align}
The kinematical function $\cS^{(III,\ab)}_{ik}$ stems from the
abelian-type term proportional to $\gab$ in Eqs.~(\ref{bfgamma}) and (\ref{gamma}), and we obtain
\begin{align}
  &\cS^{(III,\ab)}_{ik}(q_1,q_2,q_3) =\Bigg[
  \frac{1}{(q_{123}^2)^2 \per \pq{i}{123} \per \pq{k}{123}} \per
  \sg \frac{\qq{2}{3}}{\qq{1}{2} \per \qq{1}{3}} \per \pst 2 \per \pp{i}{k} \per
  \qq{2}{3}+(d-4) \per \pq{i}{1} \per \pq{k}{1}\nonumber \\
  &-2 \per \pq{i}{1} \per
  \pq{k}{23}-4 \per \pq{i}{2} \per \pq{k}{3} \pdt +\frac{1}{\qq{1}{2}} \per \pst \pp{i}{k} \per ((d-2) \per \qq{1}{3}+4 \per
  \qq{2}{3})+2 \per (4-d) \per \pq{i}{1} \per \pq{k}{2}\nonumber \\
  &+2 \per (2-d) \per \pq{i}{1} \per \pq{k}{3}+4 \per \pq{i}{2} \per
  p_k(q_{2}-q_{3})\pdt \, + (d-4) \per \pp{i}{k}\dg
  +(2\lra 3) \Bigg]+ (i\lra k) \;.
\label{sIIIab}
\end{align}
The kinematical function $\cS^{(III,\na)}_{ik}$ comes from 
the non-abelian term proportional to $\gna$ in Eqs.~(\ref{bfgamma}) and (\ref{gamma}),
and we have
\begin{align}
  &\cS^{(III,\na)}_{ik}(q_1,q_2,q_3) =\Bigg[
  \frac{1}{\pq{i}{123} \per \pq{k}{123}} \per \sg \frac{1}{(q_{123}^2)^2} \per \sq\frac{8 \per \pp{i}{k} \per \qq{1}{2} \per
    \qq{1}{3}}{(\qq{2}{3})^2}\nonumber \\
&+\frac{2 \per ((d-6) \per p_i(q_{1}-q_{3})-(2+d)
    \per \pq{i}{2}) \per \pq{k}{2} \per \qq{1}{3}}{\qq{1}{2} \per
    \qq{2}{3}}+\frac{\qq{2}{3}}{\qq{1}{2} \per \qq{1}{3}} \per (-2 \per \pp{i}{k} \per
\qq{2}{3}\nonumber\\
&+\pq{i}{1} \per ((4-d) \per \pq{k}{1}+2 \per \pq{k}{2})+2 \per
\pq{i}{2} \per (\pq{k}{1}+2 \per \pq{k}{3}))+\frac{1}{\qq{2}{3}} \per (8 \per \pp{i}{k} \per \qq{1}{3}+\pq{i}{1} \per
((2-d) \per \pq{k}{1}\nonumber\\
&+2 \per (d-6) \per \pq{k}{2})+8 \per \pq{i}{2} \per
p_k(q_{3}-q_{2}))+\frac{1}{\qq{1}{2}} \per (\pp{i}{k} \per ((d-2) \per \qq{1}{3}-4 \per
\qq{2}{3})+\pq{i}{1} \per (2 \per (6-d) \per \pq{k}{1}\nonumber\\
&+2 \per (d-4) \per
\pq{k}{2}+(10-d) \per \pq{k}{3})+2 \per \pq{i}{2} \per ((d+2) \per \pq{k}{1}-2 \per \pq{k}{2}+6 \per
\pq{k}{3})\nonumber\\
&+\pq{i}{3} \per ((d-2) \per \pq{k}{1}+8 \per \pq{k}{2}))\nonumber\\
&+(4-d) \per \pp{i}{k} \dq +\frac{1}{q_{123}^2} \per \sq\frac{1}{(\qq{2}{3})^2} \per
  \left(\left(\frac{1}{\pq{i}{1}}-\frac{1}{\pq{i}{23}}\right) \per 2 \per
    \pq{i}{2} \per (2 \per \pp{i}{k} \per \qq{1}{3}-2 \per \pq{i}{2} \per
    \pq{k}{3}-\pq{i}{3} \per \pq{k}{1})\right)\nonumber\\
&+\frac{1}{\qq{1}{2} \per \qq{2}{3}} \per
\left(\frac{1}{\pq{i}{1}}-\frac{1}{\pq{i}{23}}\right) \per ((2 \per \pp{i}{k}
\per \pq{i}{2}-m_i^2 \per \pq{k}{2}) \per \qq{1}{3}-2 \per \pq{i}{2} \per
(\pq{i}{12} \per \pq{k}{3}+\pq{i}{3} \per \pq{k}{12}))\nonumber\\
&+\frac{1}{\qq{2}{3}} \per
\left(\left(\frac{1}{\pq{i}{1}}-\frac{1}{\pq{i}{23}}\right) \per (m_i^2 \per
  p_k(q_{1}-q_{2})+2 \per \pp{i}{k} \per (2 \per \pq{i}{2}-\pq{i}{1}))\right)\nonumber\\
&+\frac{1}{\qq{1}{2}} \per
  \left(\left(\frac{1}{\pq{i}{1}}-\frac{1}{\pq{i}{23}}\right) \per (2 \per
    \pp{i}{k} \per \pq{i}{2}+m_i^2 \per \pq{k}{1})-4 \per
    \pp{i}{k}\right)\dq\nonumber\\
&+\frac{\pp{i}{k} \per \pq{i}{2} \per \pq{k}{3}}{2 \per (\qq{2}{3})^2} \per
\left(\frac{1}{\pq{i}{1}}-\frac{1}{\pq{i}{23}}\right) \per
\left(\frac{1}{\pq{k}{1}}-\frac{1}{\pq{k}{23}}\right)\nonumber\\
&+\frac{-(\pp{i}{k})^2}{4 \per \qq{2}{3}} \per
\left(\frac{1}{\pq{i}{1}}-\frac{1}{\pq{i}{23}}\right) \per
\left(\frac{1}{\pq{k}{1}}-\frac{1}{\pq{k}{23}}\right)\dg
+(2\lra 3)\Bigg] + (i\lra k)\;.
\label{sIIIna}
\end{align}

In summary, the squared current for soft $gq{\bar q}$ emission is obtained by summing 
the contributions of Eqs.~(\ref{Jsq1}), (\ref{WI}), (\ref{W2}) and (\ref{Gammasq}), and we find
\begin{align} \label{J123sq}
  |\bj(q_1,q_2,q_3)|^2 &= \sy{|\bj(q_1)|^2}{|\bj(q_2,q_3)|^2} + W(q_1,q_2,q_3) \;,
\end{align}
where
\begin{align}
  W(q_1,q_2,q_3) &= W^{(I)}(q_1,q_2,q_3) +W^{(II)}(q_1,q_2,q_3) +W^{(III)}(q_1,q_2,q_3) \nonumber \\
  &\eqcs-\left( \g\,\mu^\ep\right)^6 T_R  \;\Bigl\{ \;\sum_{i,k} \T_i\cdot \T_k \left[
    C_A\,\cS^{(A)}_{ik}(q_1,q_2,q_3) + C_F\,\cS^{(F)}_{ik}(q_1,q_2,q_3) \right] \Bigr.
  \nn \\
  & \Bigl.  + \sum_{i,k,m} \tri{ikm} \;\cS_{ikm}(q_1,q_2,q_3) \;\Bigr\} \;\;.
\label{Wqcd}
\end{align}
The kinematical function $\cS_{ikm}$ is given in Eq.~(\ref{Sikm}),
and the kinematical functions $\cS^{(A)}_{ik}$ and $\cS^{(F)}_{ik}$ of the dipole contributions to Eq.~(\ref{Wqcd}) are
\begin{align}
  \cS^{(A)}_{ik}(q_1,q_2,q_3) &= \cS^{(I)}_{ik}(q_1,q_2,q_3)+\cS^{(II)}_{ik}(q_1,q_2,q_3) \nn \\
  &+ \frac{1}{4} \left(
   \cS^{(III,\na)}_{ik}(q_1,q_2,q_3)-\cS^{(III,\ab)}_{ik}(q_1,q_2,q_3) \right) \;\;, 
 \label{SA}\\
  \cS^{(F)}_{ik}(q_1,q_2,q_3) &=\cS^{(III,\ab)}_{ik}(q_1,q_2,q_3) \;\;.
 \label{SF}
\end{align}
where $\cS^{(I)}_{ik}, \cS^{(II)}_{ik}, \cS^{(III,\ab)}$ and $\cS^{(III,\na)}_{ik}$ are
given in Eqs.~(\ref{SI}), (\ref{SII}), (\ref{sIIIab}) and (\ref{sIIIna}).

The kinematical functions $\cS^{(A)}_{ik}, \cS^{(F)}_{ik}$ and $\cS_{ikm}$ in 
Eq.~(\ref{Wqcd}) depend on the hard-parton momenta, and we recall that our results are valid
for both massless and massive hard partons (see, e.g., the dependence on $p_i^2=m_i^2$
in Eqs.~(\ref{SI}), (\ref{SII}) and (\ref{sIIIna})).

We note that the dipole kinematical functions $\cS^{(A)}_{ik}$ and $\cS^{(F)}_{ik}$
in Eqs.~(\ref{SA}) and (\ref{SF}) (i.e., $\cS^{(III,\ab)}$ and $\cS^{(III,\na)}_{ik}$
in Eqs.~(\ref{sIIIab}) and (\ref{sIIIna})) have an explicit dependence on $\ep$ through
the number $d=4 - 2\ep$ of space-time dimensions.
Such $\ep$ dependence actually derives from the fact that we use CDR with $h_g= d-2=2(1-\ep)$
spin polarization states for the on shell soft gluon with momentum $q_1$. Other versions 
of dimensional regularizations, such as dimensional reduction (DR) \cite{Siegel:1979wq}
and the four-dimensional helicity (4DH) scheme \cite{Bern:1991aq},
use $h_g=2$ spin polarization states. The result for $|\bj(q_1,q_2,q_3)|^2$ in the DR and
4DH schemes is obtained by simply setting $\ep=0$ (i.e., $d=4$) in our expressions for
$\cS^{(A)}_{ik}$ and $\cS^{(F)}_{ik}$.

In Eq.~(\ref{J123sq}) the squared current for the radiation of the three soft partons
is decomposed in terms of irreducible correlation contributions, similar to the analogous decomposition of the squared currents for double \cite{Catani:1999ss} and triple 
\cite{Catani:2019nqv} soft-gluon radiation.
The first term in the right-hand side of Eq.~(\ref{J123sq}) represents the `independent'
(though colour-symmetrized) radiation of a soft gluon and a soft-$q{\bar q}$ pair
according to the corresponding squared currents $|\bj(q_1)|^2$ (see Eq.~(\ref{j1q}))
and $|\bj(q_2,q_3)|^2$ (see Eq.~(\ref{j23q})). The term $W(q_1,q_2,q_3)$ in 
Eq.~(\ref{J123sq}) is an irreducible correlation contribution for soft $gq\bar q$ radiation.

The colour structure of $W(q_1,q_2,q_3)$ in Eq.~(\ref{Wqcd}) is given in terms of dipole
operators $\T_i\cdot \T_k$ and tripole operators $\tri{ikm}$ (here, two or three of the 
hard-parton indices $i, k$ and $m$ can also be equal). The presence in the tree-level squared current $|\bj(q_1,q_2,q_3)|^2$ of colour correlations due to the tripole operators
$\tri{ikm}$ was first observed in Ref.~\cite{DelDuca:2022noh}.
We note that the contributions to Eq.~(\ref{Wqcd}) that are proportional to 
$\cS^{(A)}_{ik}$ have a purely non-abelian origin. The other contributions to
$W$ in Eq.~(\ref{Wqcd}), namely, the dipole terms proportional to $\cS^{(F)}_{ik}$
and the tripole terms, have an `abelian character', since corresponding irreducible correlations also occur for soft photon-lepton-antilepton radiation in QED
(see Sect.~\ref{s:qed}).

The structure of Eqs.~(\ref{J123sq}) and (\ref{Wqcd}) is identical to that obtained in 
Ref.~\cite{DelDuca:2022noh}. Exploiting colour conservation and the symmetries of
$\T_i\cdot \T_k$ and $\tri{ikm}$ with respect to their parton indices, the kinematical
functions $\cS^{(A)}_{ik}, \cS^{(F)}_{ik}$ and $\cS_{ikm}$ can be written in different ways,
without affecting the value of $W(q_1,q_2,q_3)$ in Eq.~(\ref{Wqcd}) onto colour singlet states (scattering amplitudes). Our explicit expressions of these kinematical functions appear to be more compact than the related expressions presented in 
Ref.~\cite{DelDuca:2022noh}. Considering the action of $W$ onto colour singlet states, we have carried out numerical comparisons between our result and the result of 
Ref.~\cite{DelDuca:2022noh}, and we find complete agreement.

The irreducible correlation  $W(q_1,q_2,q_3)$ in Eq.~(\ref{Wqcd}) can be rewritten in the following equivalent form:
\begin{align}
  W(q_1,q_2,q_3) 
  &\eqcs-\left( \g\,\mu^\ep\right)^6 T_R  \;\Bigl\{ \;\frac{1}{2} \sum_{i \neq k}
 \T_i\cdot \T_k \left[
    C_A\,w^{(A)}_{ik}(q_1,q_2,q_3) + C_F\,w^{(F)}_{ik}(q_1,q_2,q_3) \right] \Bigr.
  \nn \\
  & \Bigl.  + \frac{1}{2} \sum_{i \neq k} \tri{iik}\;w^{(tri)}_{ik}(q_1,q_2,q_3)
+ \sum_{{\rm dist.}\{i,k,m\}} \tri{ikm} \;w^{(tri)}_{ikm}(q_1,q_2,q_3) \;\Bigr\} \;\;,
\label{Wred}
\end{align}
where the subscript `${\rm dist.}\{i,k,m\}$' in $\sum_{{\rm dist.}\{i,k,m\}}$
denotes the sum over distinct hard-parton indices $i,k$ and $m$
(i.e., $i \neq k, k\neq m, m\neq i$).
The dipole kinematical functions $w^{(A)}_{ik}$ and $w^{(F)}_{ik}$ are related to the corresponding functions $\cS^{(A)}_{ik}$ and $\cS^{(F)}_{ik}$ in Eq.~(\ref{Wqcd}),
and we have
\begin{equation}\label{w3ik}
  w^{(r)}_{ik}(q_1,q_2,q_3) = \cS^{(r)}_{ik}(q_1,q_2,q_3) + \cS^{(r)}_{ki}(q_1,q_2,q_3)
  - \cS^{(r)}_{ii}(q_1,q_2,q_3) - \cS^{(r)}_{kk}(q_1,q_2,q_3) \,, \;
(r=A,F)\,.
\end{equation}
The kinematical functions $w^{(tri)}_{ik}$ and $w^{(tri)}_{ikm}$ depend on the tripole
functions $\cS_{ikm}$ in Eq.~(\ref{Wqcd}), and we obtain
\begin{align}
  w^{(tri)}_{ik}(q_1,q_2,q_3) &= \Bigl[  \cS_{iik}(q_1,q_2,q_3) + \cS_{iki}(q_1,q_2,q_3) 
+  \cS_{kii}(q_1,q_2,q_3) - \cS_{iii}(q_1,q_2,q_3) \Bigr] \nn \\
& - \;( i \leftrightarrow k) \;\;,
\label{wiik}
\end{align}
\begin{align}
  w^{(tri)}_{ikm}(q_1,q_2,q_3) &= \cS_{ikm}(q_1,q_2,q_3) - \frac{1}{2} \Bigl[  \cS_{iik}(q_1,q_2,q_3) + \cS_{iki}(q_1,q_2,q_3) \Bigr. \nn \\ 
& + \Bigl. \cS_{kii}(q_1,q_2,q_3) - \cS_{iii}(q_1,q_2,q_3) \Bigr] \;\;.
\label{wikm}
\end{align}
The equality of Eqs.~(\ref{Wqcd}) and (\ref{Wred}) follows from colour conservation and the symmetries of the operators  $\T_i\cdot \T_k$ and $\tri{ikm}$ with respect to their 
hard-parton indices. The proof of the equivalence between the two expressions in 
Eqs.~(\ref{Wqcd}) and (\ref{Wred}) is presented at the end of this section.

The expression in Eq.~(\ref{Wred}) involves colour correlations between two and three
distinct hard partons. The size of the two hard-parton correlations is controlled by the kinematical functions $w^{(A)}_{ik}, w^{(F)}_{ik}$ and $w^{(tri)}_{ik}$, which are
are physically related to the interaction of two hard partons, $i$ and $ k$, in a colour singlet configuration (see Sect.~\ref{s:e2h}).

The structure of $W(q_1,q_2,q_3)$ in Eq.~(\ref{Wred}) is almost identical to that of the 
one-loop QCD corrections to the squared current for soft-$q{\bar q}$ radiation
(see Eq.~(61) in Ref.~\cite{Catani:2021kcy}). The only difference is due to the presence of three-parton correlations of the type $f^{abc} T^a_i T^b_k T^c_m$ in the case of the one-loop
squared current. Such correlations have a one-loop absorptive origin \cite{Catani:2021kcy}
and, therefore, they are absent in the tree-level $gq{\bar q}$ correlation term 
$W(q_1,q_2,q_3)$. In Ref.~\cite{Catani:2021kcy} the tripole operator with two distinct
indices is written as $\tri{iik}= {\widetilde{\bf D}}_i \cdot \T_k = \sum_a D^a_i T^a_k$
in terms of the `$d$-conjugated' charge operator  
$D^a_i = \sum_{b,c} d^{abc} \,T^b_i \,T^c_i$.
We note that the kinematical function  $w^{(tri)}_{ik}$ in Eq.~(\ref{wiik}) is antisymmetric under the exchange $i \leftrightarrow k$ of the hard-parton indices.
Such antisymmetry of $w^{(tri)}_{ik}$ implies that in the sum over $i$ and $k$ of 
Eq.~(\ref{Wred}) we can replace $\tri{iik}$ by its antisymmetric component, namely,
$\tri{iik} \to (\tri{iik}- \tri{kki})/2$.

Some main features of $W$ in Eq.~(\ref{Wred}) (or, equivalently, Eq.~(\ref{Wqcd}))
are fully similar to those of the analogous one-loop corrections in 
Ref.~\cite{Catani:2021kcy}. The dipole kinematical functions $w^{(r)}_{ik}(q_1,q_2,q_3)$
($r=A,F$) in Eq.~(\ref{Wred}) (and $\cS^{(r)}_{ik}(q_1,q_2,q_3)$ in Eq.~(\ref{Wqcd}))
are {\it symmetric} with respect to the exchange $q_2 \leftrightarrow q_3$ of the quark and antiquark momenta. In contrast, the functions $w^{(tri)}_{ik}(q_1,q_2,q_3)$ and
$w^{(tri)}_{ikm}(q_1,q_2,q_3)$ in Eq.~(\ref{Wred}) (and $\cS_{ikm}(q_1,q_2,q_3)$ in
Eq.~(\ref{Wqcd})) are {\it antisymmetric} with respect to the exchange $q_2 \leftrightarrow q_3$ and, therefore, they produce a quark--antiquark {\it charge asymmetry} in the tree-level
squared current $|\bj(q_1,q_2,q_3)|^2$. We note that such charge asymmetry functions contribute to $W(q_1,q_2,q_3)$ with the associated colour factors $\tri{iik}$
and $\tri{ikm}$ that have a linear dependence on the colour tensor $d^{abc}$. Therefore,
since $d^{abc}$ is odd under charge conjugation, the charge asymmetry features of 
$|\bj(q_1,q_2,q_3)|^2$ are consistent with the charge conjugation invariance of the QCD interactions (see also related comments in Sect.~\ref{s:e2h} and \ref{s:e3h}
and in Ref.~\cite{Catani:2021kcy}).
In particular, we note \cite{Catani:2021kcy} that the charge asymmetry contributions of
$|\bj(q_1,q_2,q_3)|^2$ vanish if the squared current acts on a pure multigluon scattering
amplitude $\M(\{p_i\})$, namely, if $\M(\{p_i\})$ has only gluon external lines (with no additional
external $q{\bar q}$ pairs or colourless particles). We also note \cite{Catani:2021kcy}
that the three-particle correlations of the type $\tri{ikm}$ with three distinct partons
contribute only to processes with {\it four} or {\it more} hard partons.
General properties of the colour algebra of the $d$-type tripoles $\tri{ikm}$
and their action onto two and three hard-parton states are discussed in 
Ref.~\cite{Catani:2021kcy} (see also Appendix~\ref{a:tthp}).

We conclude this section by proving the equivalence of Eqs.~(\ref{Wqcd}) and (\ref{Wred}).
In the case of the dipole contributions proportional to $\T_i\cdot \T_k$, the equivalence directly follows from the colour conservation relation in Eq.~(\ref{colcons}),
in the same way as the equivalence between Eqs.~(\ref{J1sq})--(\ref{I23})
and Eqs.~(\ref{j1q})--(\ref{w23}). Considering the tripole contributions to 
Eq.~(\ref{Wqcd}), we can write
\begin{equation}
\label{sumtri}
\sum_{i,k,m} \tri{ikm} \,\cS_{ikm} = \sum_i \tri{iii} \,\cS_{iii}
+ \sum_{i \neq k}  \tri{iik} \,\left( \cS_{iik} +  \cS_{iki} + \cS_{kii}\right) +
\sum_{{\rm dist.}\{i,k,m\}} \tri{ikm} \,\cS_{ikm} \;,
\end{equation}
where we have separated the terms with three equal parton indices, two equal indices
and three distinct indices, and we have also used the symmetry property
$\tri{iki}=\tri{kii}=\tri{iik}$.
Considering the contributions to Eq.~(\ref{Wred}) with three distinct parton indices,
we can write
\begin{equation}
\label{sumtridist}
\sum_{{\rm dist.}\{i,k,m\}} \tri{ikm} \,w^{(tri)}_{ikm} \; \eqcs \;
\sum_{{\rm dist.}\{i,k,m\}} \tri{ikm} \,\cS_{ikm} 
+ \frac{1}{2} \sum_{i \neq k} \left( \tri{iki} + \tri{ikk} \right)
\left( \cS_{iik} +  \cS_{iki} + \cS_{kii} - \cS_{iii}\right) \,,
\end{equation}
where we have used the expression of $w^{(tri)}_{ikm}$ in Eq.~(\ref{wikm}) and we have performed the sum over $m$ in the terms of $w^{(tri)}_{ikm}$ that do not depend on $m$. Exploiting colour conservation, we have used the relation
$\sum_{m\neq i,k} \tri{ikm} \,\eqcs\, - (\tri{iki} + \tri{ikk})$ to explicitly
carry out the sum over $m$. We can now consider the difference between the terms
in the left-hand side of Eqs.~(\ref{sumtri}) and (\ref{sumtridist}), and we obtain
\begin{align}
\!\!
\sum_{i,k,m} \tri{ikm} \,\cS_{ikm} - \sum_{{\rm dist.}\{i,k,m\}} \tri{ikm} \,w^{(tri)}_{ikm} 
\, &\eqcs \, \frac{1}{2} \sum_{i \neq k} \left( \tri{iik} - \tri{ikk} \right)
\left( \cS_{iik} +  \cS_{iki} + \cS_{kii} - \cS_{iii}\right) 
\label{difftri} \\
&=  \sum_{i \neq k} \,\tri{iik} \;w^{(tri)}_{ik} \;\;.
\label{difffin}
\end{align}
In Eq.~(\ref{difftri}) we have first inserted the expressions in the right-hand side of 
Eqs.~(\ref{sumtri}) and (\ref{sumtridist}), and then we have used 
$\tri{iki}=\tri{iik}$ and the relation $\tri{iii} \eqcs - \sum_{k \neq i} \tri{iik}$
(which follows from colour conservation). In Eq.~(\ref{difffin}) we have first renamed 
$i \leftrightarrow k$ in the terms of Eq.~(\ref{difftri}) that are proportional to
$\tri{ikk}$, and then we have simply used the expression of $w^{(tri)}_{ik}$ in 
Eq.~(\ref{wiik}). In conclusions, Eq.~(\ref{difffin}) proves the equality of the charge asymmetry contributions in Eqs.~(\ref{Wqcd}) and (\ref{Wred}).

\section{Processes with two and three hard partons\label{s:e23h}}

The simplest applications of the QCD soft-factorization formula (\ref{softsquared})
regard processes with two and three hard partons.
In these processes the structure of the colour correlations produced by soft emission is
simplified \cite{Catani:1999ss, Catani:2019nqv, Catani:2021kcy}.
In this section we present the explicit expressions of $| \J(q_1,q_2,q_3) |^2$
for $gq{\bar q}$ emission from amplitudes with two and three hard partons and, 
in particular, we highlight the corresponding charge asymmetry effects.

\subsection{Soft $gq{\bar q}$ emission from two hard partons\label{s:e2h}}

We consider a generic scattering amplitude $\M_{BC}(\{p_i\})$ whose
external legs are two hard partons (denoted as $B$ and $C$), with momenta $p_B$ and $p_C$,
and additional colourless particles.
The two hard partons can be either a $q{\bar q}$
pair ($\{ BC \} = \{ q{\bar q} \}$) (note the we specify $B=q$ and $C={\bar q}$)
or two gluons ($\{ BC \} = \{ gg \}$). There is only {\em one} colour singlet configuration
of the two hard partons, and the corresponding one-dimensional colour space is generated by a single colour state vector that we denote as $\ket{B C}$. The colour space amplitude
$\ket{\M_{BC}}$ is a colour singlet state and, therefore, it is directly proportional to 
$\ket{B C}$.
 
The squared current $|\J(q_1,\cdots,q_N) |^2$ in Eq.~(\ref{softsquared})
conserves the colour charge of the
hard partons and, consequently, the state $|\J|^2 \,\ket{B C}$ is also
proportional to $\ket{B C}$ and we have
\beq
\label{jbc}
| \J(q_1,\cdots,q_N) |^2 \;\,\ket{B C} = \ket{B C} 
\;| \J(q_1,\cdots,q_N) |^{2}_{\; BC} \;,
\eeq
\beq
\label{softsquaredAB}
\bra{\M_{BC} (\{p_i\})} \;| \J(q_1,\cdots,q_N) |^2 \;\ket{\M_{BC} (\{p_i\})} 
= | \M_{BC} (\{p_i\}) |^2 \;\, | \J(q_1,\cdots,q_N) |^{2}_{\; BC} \;,
\eeq
where $| \J|^{2}_{\; BC}$ is a $c$-number (it is the eigenvalue of the operator $| \J|^{2}$
onto the colour state $\ket{B C}$).

We note that the right-hand side of Eq.~(\ref{softsquaredAB}) is proportional to the squared amplitude $| \M_{BC} (\{p_i\}) |^2$ with no residual colour correlations between the hard partons $B$ and $C$. In this respect, the structure of soft factorization in 
Eq.~(\ref{softsquaredAB}) is similar to that of soft-photon factorization in QED.

We recall \cite{Catani:2019nqv} that Eqs.~(\ref{jbc}) and (\ref{softsquaredAB})
are valid for squared currents $| \J(q_1,\cdots,q_N) |^2$ of an arbitrary number $N$ and an arbitrary type (gluons and quark--antiquark pairs) of soft partons. 
We also recall \cite{Catani:2019nqv}  that Eqs.~(\ref{jbc}) and (\ref{softsquaredAB})
are valid at {\it arbitrary} loop orders in the perturbative expansion
of both the squared current and the squared amplitude.

We evaluate $| \J(q_1,q_2,q_3) |^{2}_{\; BC}$ for soft $gq{\bar q}$ emission at the tree level by using Eq.~(\ref{J123sq}). The squared current terms 
$| \J(q_1) |^{2}_{\; BC}$ for single-gluon emission and $| \J(q_2,q_3) |^{2}_{\; BC}$
for soft-$q{\bar q}$ emission are well known \cite{Catani:1999ss}. The correlation term 
$W(q_1,q_2,q_3)$ of Eq.~(\ref{J123sq}) depends on dipole and tripole colour operators
(see Eqs.~(\ref{Wqcd}) or (\ref{Wred})). The action of the dipole operators onto
$\ket{B C}$ is elementary \cite{csdip},
and the action of the tripole operators is explicitly evaluated in 
Ref.~\cite{Catani:2021kcy} (see also Appendix~\ref{a:tthp}). We straightforwardly obtain the following result:
\begin{align}
\!\!  | \J(q_1,q_2,q_3) |^{2}_{\; BC}&= \left( \g\,\mu^\ep\right)^6 T_R \,C_F
  \Bigl\{ \,\Bigl[ C_F  \left( w_{BC}(q_1) \;w_{BC}(q_2,q_3) + w^{(F)}_{BC}(q_1,q_2,q_3) \right) \Bigr. \Bigr. \nn \\
& + \Bigl. \Bigl. C_A \;w^{(A)}_{BC}(q_1,q_2,q_3) \Bigr] 
+ \frac{1}{2} \,d_A \,w^{(tri)}_{BC}(q_1,q_2,q_3)\Bigr\} \;\;, \;\;
( \{B=q, C={\bar q}\}) \;, 
\label{BCqq}\\
\!\!  | \J(q_1,q_2,q_3) |^{2}_{\; BC}&=\left( \g\,\mu^\ep\right)^6 T_R \,C_A
  \,\Bigl\{ \; C_A  \left( w_{BC}(q_1) \;w_{BC}(q_2,q_3) + w^{(A)}_{BC}(q_1,q_2,q_3) \right) \Bigr.  \nn \\
& +  C_F \;w^{(F)}_{BC}(q_1,q_2,q_3) 
\Bigr\} \;\;, \quad \quad \quad \quad \quad \quad \quad \quad \quad \quad \quad 
 ( \{B C\}=\{gg\}) \;,
\label{BCgg}
\end{align}
where the colour coefficient $d_A$ is related to $d^{abc}$ as follows
\begin{equation}
\sum_{bc} \,d^{abc} \,d^{dbc} = d_A \;\delta^{ad} \;\;, \quad\quad d_A= \frac{N_c^2-4}{N_c} \;\;.
\end{equation}
The dipole kinematical functions 
$w_{BC}(q_1), w_{BC}(q_2,q_3),  w^{(A)}_{BC}(q_1,q_2,q_3)$ and $w^{(F)}_{BC}(q_1,q_2,q_3)$
are given in Eqs.~(\ref{w1}), (\ref{w23}) and (\ref{w3ik}),
and they are {\it symmetric} under the exchange $B \leftrightarrow C$ of the hard partons
(i.e., the exchange $p_B \leftrightarrow p_C$ of the hard-parton momenta).
The tripole kinematical function $w^{(tri)}_{BC}(q_1,q_2,q_3)$ is given in Eq.~(\ref{wiik}),
and it is {\it antisymmetric} under the exchange $B \leftrightarrow C$.
The term that is proportional to $w_{BC}(q_1) w_{BC}(q_2,q_3)$ in 
Eqs.~(\ref{BCqq}) and (\ref{BCgg}) is due to the independent emission of the soft gluon
and of the soft $q{\bar q}$ pair. All the other contributions to 
Eqs.~(\ref{BCqq}) and (\ref{BCgg}) are due to irreducible correlations for soft $gq{\bar q}$
emission.

In the case of soft $gq{\bar q}$ emission form the hard partons $\{BC\}=\{q{\bar q}\}$,
the square-bracket contribution in the right-hand side of Eq.~(\ref{BCqq}) is 
symetric with respect to the exchange $q_2 \leftrightarrow q_3$ of the momenta of the soft
quark and antiquark. The kinematical function $w^{(tri)}_{BC}(q_1,q_2,q_3)$ is instead
antisymmetric under the exchange $q_2 \leftrightarrow q_3$ and, therefore, the result
in Eq.~(\ref{BCqq}) explicitly shows the presence of charge asymmetry effects.
Since the function $w^{(tri)}_{BC}(q_1,q_2,q_3)$ is antisymmetric with respect to the separate exchanges $q_2 \leftrightarrow q_3$ and $B \leftrightarrow C$, in Eq.~(\ref{BCqq})
the asymmetry in the momenta of the soft quark and antiquark is correlated with a corresponding asymmetry in the momenta $p_B$ and $p_C$ of the hard quark and antiquark.
In particular, $| \J(q_1,q_2,q_3) |^{2}_{\; BC}$ is invariant under the overall
exchange of fermions and antifermions (i.e., $\{q_2, p_B \} \leftrightarrow \{q_3, p_C \}$),
consistently with charge conjugation invariance.

In the case of soft $gq{\bar q}$ radiation form two hard gluons, the tree-level result
in Eq.~(\ref{BCgg}) shows no charge asymmetry effects. As argued in 
Ref.~\cite{Catani:2021kcy}
on the basis of charge conjugation invariance, the absence of charge asymmetry effects
in $| \J(q_1,q_2,q_3) |^{2}_{\; BC}$ for $\{BC\}=\{gg\}$ persists at arbitrary orders in the QCD loop expansion.

\subsection{Soft $gq{\bar q}$ emission from three hard partons\label{s:e3h}}

We consider a generic scattering amplitude $\M_{ABC}(\{p_i\})$ whose
external legs are three hard partons and additional colourless particles.
The three hard partons, which are denoted as $A, B, C$ (with momenta $p_A, p_B ,p_C$),
can be either a gluon and a $q{\bar q}$ pair ($\{A B C\}=\{g q {\bar q}\}$)
or three gluons ($\{A B C\}=\{g g g\}$). If $\{A B C\}=\{g q {\bar q}\}$, there is only one colour singlet state that can be formed by the three hard partons.
If the three hard partons $\{A B C\}$ are gluons, they can form two distinct colour singlet states. The different dimensionality of the colour singlet space for the two cases leads
to different features of the associated soft radiation. We discuss the cases 
$\{A B C\}=\{g q {\bar q}\}$ and $\{A B C\}=\{g g g\}$ in turn.

In the case $\{A B C\}=\{g q {\bar q}\}$, we specifically set $A=g$, $B=q$ and $C={\bar q}$.
The one-dimensional colour singlet space of the three hard partons is generated by the
state vector $\ket{A B C}$, and the colour space amplitude
$\ket{\M_{ABC}}$ is directly proportional to 
$\ket{A B C}$. Since we are dealing with a one-dimensional colour singlet space, we can use the same reasoning as in Sect.~\ref{s:e2h}. 
The state $\ket{A B C}$ is an eigenstate of the squared current $|\J(q_1,\cdots,q_N) |^2$
in Eq.~(\ref{softsquared}), and we have
\beq
\label{jgqq}
| \J(q_1,\cdots,q_N) |^2 \;\,\ket{A B C} = \ket{ A B C} 
\;| \J(q_1,\cdots,q_N) |^{2}_{\; ABC} \;, \quad \quad (\{A B C\}=\{g q {\bar q}\}) \;,
\eeq
\beq
\label{sgqq}
\bra{\M_{ABC} (\{p_i\})} \;| \J(q_1,\cdots,q_N) |^2 \;\ket{\M_{ABC} (\{p_i\})} 
= | \M_{ABC} (\{p_i\}) |^2 \;\, | \J(q_1,\cdots,q_N) |^{2}_{\; ABC} \;,
\eeq
where $| \J|^{2}_{\; ABC}$ is a $c$-number.

Similar to Eqs.~(\ref{jbc}) and (\ref{softsquaredAB}), we recall \cite{Catani:2019nqv} 
that Eqs.~(\ref{jgqq}) and (\ref{sgqq}) are valid for {\em arbitrary} squared
currents $| \J(q_1,\cdots,q_N) |^2$ and at {\em arbitrary} loop orders.
 
Considering the tree-level squared current $| \J(q_1,q_2,q_3) |^{2}$ for soft
$g g {\bar q}$ emission, the eigenvalue $| \J(q_1,q_2,q_3) |^{2}_{\; ABC}$ in 
Eqs.~(\ref{jgqq}) and (\ref{sgqq}) can be written in the following form:
\begin{align}
| \J(q_1,q_2,q_3) |^{2}_{\; ABC} &= \left( \g\,\mu^\ep\right)^6 \,T_R
\Bigl[ \, F_{ABC}^{({\rm in.em.})}(q_1,q_2,q_3) 
+ W_{ABC}^{({\rm ch.sym.})}(q_1,q_2,q_3) \Bigr. \nn \\
&+ \Bigl. W_{ABC}^{({\rm ch.asym.})}(q_1,q_2,q_3) \,\Bigr] \;, 
\quad \quad \quad \quad (\{A B C\}=\{g q {\bar q}\}) \;,
\label{softgqq}
\end{align}
which directly derives from Eqs.~(\ref{J123sq}) and (\ref{Wred}).

The functions $F_{ABC}^{({\rm in.em.})}$ and $W_{ABC}^{({\rm ch.sym.})}$ 
in Eq.~(\ref{softgqq}) are
\begin{align}
F_{ABC}^{({\rm in.em.})}(q_1,q_2,q_3) &= 
\bigl[ C_F \;w_{BC}(q_1) + C_A \;w_{ABC}(q_1) \bigr] \nn \\
& \times \bigl[ C_F \;w_{BC}(q_2,q_3) + C_A \;w_{ABC}(q_2,q_3) \bigr] \;,
\quad \;\; (\{A B C\}=\{g q {\bar q}\}) \;,
\label{fgqq}
\end{align}
\begin{align}
W_{ABC}^{({\rm ch.sym.})}(q_1,q_2,q_3) &= C_F^2 \;w_{BC}^{(F)}(q_1,q_2,q_3)  
  + C_F C_A \left( w_{BC}^{(A)}(q_1,q_2,q_3) + w_{ABC}^{(F)}(q_1,q_2,q_3) \right)  \nn \\
  &+ C_A^2 \;w_{ABC}^{(A)}(q_1,q_2,q_3) \;,
\quad \quad \quad \quad \quad \quad \quad  \;\;\;\;\; (\{A B C\}=\{g q {\bar q}\}) \;,
\label{csgqq}
\end{align}
where the two hard-parton functions
$w_{ik}(q_1), w_{ik}(q_2,q_3),  w^{(r)}_{ik}(q_1,q_2,q_3) \,(r=F,A)$ 
are given in Eqs.~(\ref{w1}), (\ref{w23}), (\ref{w3ik}),
and we have used them to define the corresponding three hard-parton functions 
$w_{ABC}^{(r)}$\footnote{In the cases of one and two soft momenta $(N=1,2)$, the explicit superscripts $(r)$ have to be removed in Eq.~(\ref{wabc}).}:
\begin{equation}\label{wabc}
  w_{ABC}^{(r)}(q_1,\cdots,q_N) \equiv 
  \frac{1}{2} \left[ w_{AB}^{(r)}(q_1,\cdots,q_N) + 
    w_{AC}^{(r)}(q_1,\cdots,q_N) - w_{BC}^{(r)}(q_1,\cdots,q_N) 
    \right] \;.
\end{equation}
The function $W_{ABC}^{({\rm ch.asym.})}$ in Eq.~(\ref{softgqq}) is
\begin{align}
W_{ABC}^{({\rm ch.asym.})}(q_1,q_2,q_3) &= \frac{d_A}{4}
 \Bigl[  
- C_A \, \left(w^{(tri)}_{BC}(q_1,q_2,q_3) + w^{(tri)}_{CA}(q_1,q_2,q_3) 
    + w^{(tri)}_{AB}(q_1,q_2,q_3) \right)
\Bigr.  \nn \\
&\Bigl. + \,2 C_F \;w^{(tri)}_{BC}(q_1,q_2,q_3) 
    \Bigr] \;\;, 
\quad \quad \quad \quad \quad \;\;\;\;\;\; (\{A B C\}=\{g q {\bar q}\}) \;,
\label{cagqq}
\end{align}
where the kinematical function $w^{(tri)}_{ik}$ is given in Eq.~(\ref{wiik}).
As discussed below, the symmetry properties of the kinematical functions
$w_{ik}(q_1), w_{ik}(q_2,q_3),  w^{(r)}_{ik}(q_1,q_2,q_3)$ and 
$w^{(tri)}_{ik}(q_1,q_2,q_3)$ with respect to their dependence 
on the hard and soft momenta (see Sect.~\ref{s:gqqsc})
lead to ensuing symmetry properties of the functions $F^{({\rm in.em.})}, 
W^{({\rm ch.sym.})}$ and $W^{({\rm ch.asym.})}$.

The term $F^{({\rm in.em.})}_{ABC}$ in Eq.~(\ref{softgqq}) is the contribution due to the independent emission of the soft gluon and the soft $q{\bar q}$ pair. It originates from the
term $\sy{|\bj(q_1)|^2}{|\bj(q_2,q_3)|^2}$ in Eq.~(\ref{J123sq}). We have used the results of the action of both $|\bj(q_1)|^2$ and $|\bj(q_2,q_3)|^2$ onto the three hard-parton state
$\ket{ABC}$ \cite{Catani:1999ss}. Such squared currents only depend on colour dipole operators, whose action onto $\ket{ABC}$ is simply given in terms of casimir coeficients 
$C_F$ and $C_A$ (see, e.g., Ref.~\cite{csdip}). 
The colour dipole contributions of Eq.~(\ref{Wred}) to the correlation term
$W(q_1,q_2,q_3)$ of the squared current in Eq.~(\ref{J123sq}) produce the corresponding
irreducible correlation contribution $W^{({\rm ch.sym.})}_{ABC}$ in Eq.~(\ref{softgqq}).
We note that both $F^{({\rm in.em.})}_{ABC}(q_1,q_2,q_3)$ and 
$W^{({\rm ch.sym.})}_{ABC}(q_1,q_2,q_3)$ are symmetric under the exchange $q_2 \leftrightarrow q_3$ (see Eqs.~(\ref{fgqq}) and (\ref{csgqq})) and, therefore, they do not lead to any charge asymmetry of the soft quark and antiquark in 
$| \J(q_1,q_2,q_3) |^{2}_{\; ABC}$. Both functions $F^{({\rm in.em.})}_{ABC}$ and
$W^{({\rm ch.sym.})}_{ABC}$ are also symmetry under the exchange $p_B \leftrightarrow p_C$
of the momenta of the hard quark and antiquark, consistently with the charge conjugation
invariance of  $| \J(q_1,q_2,q_3) |^{2}_{\; ABC}$.

The correlation term $W(q_1,q_2,q_3)$ of Eq.~(\ref{J123sq}) also includes charge asymmetry contributions. In Eq.~(\ref{Wred}) these contributions are proportional to the tripole
operators $\tri{iik}$ and $\tri{ikm}$. The action of these tripole operators onto the state
$\ket{ABC}$ of the three hard partons $\{A B C\}=\{g q {\bar q}\}$ was evaluated in 
Ref.~\cite{Catani:2021kcy} (in particular, the operator $\tri{ABC}$ with three distinct indices vanishes). Using the colour algebra results of Ref.~\cite{Catani:2021kcy}
(see also Appendix~\ref{a:tthp}), we have computed the charge asymmetry contribution to 
$| \J(q_1,q_2,q_3) |^{2}_{\; ABC}$, which is given by the function 
$W_{ABC}^{({\rm ch.asym.})}(q_1,q_2,q_3)$ in Eq.~(\ref{softgqq}).
We note the the expression of  $W_{ABC}^{({\rm ch.asym.})}$ in Eq.~(\ref{cagqq})
is antisymmetric under the exchange $p_B \leftrightarrow p_C$ of the momenta of the hard quark and antiquark, in complete analogy with the charge asymmetry contribution to 
Eq.~(\ref{BCqq}),
and consistently with the charge conjugation invariance of 
$| \J(q_1,q_2,q_3) |^{2}_{\; ABC}$ in Eq.~(\ref{softgqq}).

We now consider the case $\{A B C\}=\{g g g \}$. The three hard gluons generate a 
two-dimensional colour singlet space. We choose the basis that is formed by the two
colour state vectors $\ket{(ABC)_f \,}$ and
$\ket{(ABC)_d \,}$, which are defined as follows
\beq
\label{ABCbasis}
\bra{\,abc\,} \left(ABC\right)_f \,\rangle \equiv i f^{abc}
\,,
\;\;\;
\bra{\,abc\,} \left(ABC\right)_d \,\rangle \equiv  d^{abc}
\,,
\;\;\; \quad (\{ ABC \} = \{ ggg \}) \;\;,
\eeq
where $a, b, c$ are the colour indices of the three gluons. We note that the two states
in Eq.~(\ref{ABCbasis}) are orthogonal and have different charge conjugation.
The scattering amplitude $\ket{\cm_{ABC}(\{p_i\})}$ is, in general, a linear combination 
of the two states in Eq.~(\ref{ABCbasis}), and the action of the squared current 
$| \bj(q_1, \cdots, q_N) |^2 $ for soft-parton radiation onto $\ket{\cm_{ABC}(\{p_i\})}$
can produce colour correlations between these two states.
In general, $| \bj(q_1, \cdots, q_N) |^2 $
can be represented as a $2 \times 2$ correlation matrix that
acts onto the two-dimensional space generated by $\ket{\left( ABC \right)_f}$ and $\ket{\left( ABC \right)_d}$.
The structure of this correlation matrix is discussed in Refs.~\cite{Catani:2019nqv}
and \cite{Catani:2021kcy}
for the cases of multiple soft-gluon radiation and soft-$q{\bar q}$ radiation,
respectively. Soft $gq{\bar q}$ radiation is discussed in the following.

The action of the tree-level squared current $| \J(q_1,q_2,q_3) |^{2}$ for 
soft $gq{\bar q}$ emission onto the colour singlet states in Eq.~(\ref{ABCbasis})
can be written in the following form:
\begin{align}
| \J(q_1,q_2,q_3) |^{2} \;\ket{ABC} &= \left( \g\,\mu^\ep\right)^6 \,T_R
\Bigl[ \, F_{ABC}^{({\rm in.em.})}(q_1,q_2,q_3) 
+ W_{ABC}^{({\rm ch.sym.})}(q_1,q_2,q_3) \Bigr. \nn \\
&+ \Bigl. W_{ABC}^{({\rm ch.asym.})}(q_1,q_2,q_3) \,\Bigr] \;\ket{ABC}\;, 
\quad \quad \quad \quad (\{A B C\}=\{ggg\}) \;,
\label{J2ggg}
\end{align}
where
\begin{equation}
\label{inem3g}
F_{ABC}^{({\rm in.em.})}(q_1,q_2,q_3) = 4 C_A^2 \;E_{ABC}(q_1) \;E_{ABC}(q_2,q_3) \;\;,
\quad \quad (\{A B C\}=\{ggg\}) \;,
\end{equation}
\begin{equation}
\label{symggg}
W_{ABC}^{({\rm ch.sym.})}(q_1,q_2,q_3) = 2 C_A 
\!\left[ C_F \,E_{ABC}^{(F)}(q_1,q_2,q_3) +  C_A \,E_{ABC}^{(A)}(q_1,q_2,q_3) \right],
 (\{A B C\}=\{ggg\}) \;,
\end{equation}
\begin{equation}
\label{asymggg}
W_{ABC}^{({\rm ch.asym.})}(q_1,q_2,q_3) = 4 \;\tri{BBA} \;E_{ABC}^{(tri)}(q_1,q_2,q_3) \;\;,
\quad \quad \;\;\;\; (\{A B C\}=\{ggg\}) \;.
\end{equation}
The three hard-parton functions $E_{ABC}(q_1), E_{ABC}(q_2,q_3)$ and 
$E_{ABC}^{(r)}(q_1,q_2,q_3)$ (with $r=F,A,tri$) in Eqs.~(\ref{inem3g})--(\ref{asymggg})
are given as follows\footnote{In the cases of one and two soft momenta $(N=1,2)$, the explicit superscripts $(r)$ have to be removed in Eq.~(\ref{eabc}).}
\begin{equation}
\label{eabc}
  E_{ABC}^{(r)}(q_1,\cdots,q_N) \equiv 
  \frac{1}{4} \left[ w_{AB}^{(r)}(q_1,\cdots,q_N) + 
    w_{BC}^{(r)}(q_1,\cdots,q_N) + w_{CA}^{(r)}(q_1,\cdots,q_N) 
    \right] \;,
\end{equation}
in terms of the corresponding two hard-parton functions
$w_{ik}(q_1), w_{ik}(q_2,q_3),  w^{(r)}_{ik}(q_1,q_2,q_3)$ 
in Eqs.~(\ref{w1}), (\ref{w23}), (\ref{w3ik}) and (\ref{wiik}).
The symmetry properties of the functions $E_{ABC}^{(r)}$
are the consequence
of the corresponding symmetries of the functions 
$w_{ik}^{(r)}$
in the right-hand side of Eq.~(\ref{eabc}).
The function $E_{ABC}^{(tri)}(q_1,q_2,q_3)$ is antisymmetric under the exchange 
$q_2 \leftrightarrow q_3$ of the momenta of the soft quark and antiquark, and it is also antisymmetric under the exchange of the momenta of two hard gluons (e.g., 
$p_A \leftrightarrow p_B$). The functions $E_{ABC}^{(r)}$ in Eqs.~(\ref{inem3g}) and 
(\ref{symggg}) are instead symmetric under the exchange 
$q_2 \leftrightarrow q_3$ and have a fully symmetric dependence on the hard-gluon momenta
$p_A, p_B, p_C$.

The result in Eq.~(\ref{J2ggg}) directly derives from Eqs.~(\ref{J123sq}) and (\ref{Wred}),
and it has a structure that follows the structure of  
Eq.~(\ref{softgqq}).
The term $F^{({\rm in.em.})}_{ABC}(q_1,q_2,q_3)$ in Eq.~(\ref{J2ggg}) is the contribution
of the independent emission of the soft gluon and the soft $q{\bar q}$ pair. 
The irreducible correlation term $W^{({\rm ch.sym.})}_{ABC}(q_1,q_2,q_3)$
in Eq.~(\ref{J2ggg}) is due to the colour dipole contributions of Eq.~(\ref{Wred}) to 
$W(q_1,q_2,q_3)$. Both terms $F^{({\rm in.em.})}_{ABC}$ and $W^{({\rm ch.sym.})}_{ABC}$
originate from colour dipole interactions, whose action onto a generic hard-parton state
$\ket{ABC}$ are simply proportional to the unit operator in colour space 
\cite{Catani:1999ss}.

The term $W_{ABC}^{({\rm ch.asym.})}$ in Eq.~(\ref{J2ggg}) is due to the colour tripole
contributions of Eq.~(\ref{Wred}) to the irreducible correlation operator $W(q_1,q_2,q_3)$
in Eq.~(\ref{J123sq}). The action of the tripole operators onto the three-gluon states in Eq.~(\ref{ABCbasis}) was explicitly evaluated in Ref.~\cite{Catani:2021kcy} (see also Appendix~\ref{a:tthp}). In particular, the tripoles $\tri{ABC}$ with three distinct gluons vanish, while the tripoles with two distinct gluons are proportional to one another
(the proportionality factors are $\pm 1$). It turns out that the term $W_{ABC}^{({\rm ch.asym.})}$ in Eq.~(\ref{J2ggg}) is directly proportional to a single tripole operator
(e.g., $\tri{BBA}$) as shown in Eq.~(\ref{asymggg}). The tripole operators are odd under charge conjugation and, therefore, they act differently onto the two colour states in 
Eq.~(\ref{ABCbasis}). Considering the operator $\tri{BBA}$ in Eq.~(\ref{asymggg}), we have
\cite{Catani:2021kcy}
\beq
\label{triggg}
\tri{BBA} \,\ket{\left( ABC \right)_f} =  \frac{C_A^2}{4}  \;\ket{\left( ABC \right)_d} \;,
\;\;\;\;
\tri{BBA} \,\ket{\left( ABC \right)_d} = \frac{C_A \,d_A}{4}  \;\ket{\left( ABC \right)_f} \;,
\eeq
and we note that the tripole operators produce 
`pure' 
transitions between the colour 
symmetric and colour antisymmetric states $\ket{\left( ABC \right)_f}$ and $\ket{\left( ABC \right)_d}$,
which have different charge conjugation.

The results in Eqs.~(\ref{J2ggg})--(\ref{asymggg}) can be used to explicitly evaluate the action of the tree-level squared current $| \J(q_1,q_2,q_3) |^{2}$ onto a scattering amplitude $\ket{\cm_{ABC}(\{p_i\})}$ with three hard gluons.
The 
scattering amplitude $\ket{\cm_{ABC}(\{p_i\})}$ is, in general, a linear combination 
of the two colour states in Eq.~(\ref{ABCbasis}), and we write
\beq
\label{3gamp}
\ket{\cm_{ABC}(\{p_i\})} = \ket{\left( ABC \right)_f} \;\;\cm_{f}(p_A,p_B,p_C) \,+
\ket{\left( ABC \right)_d} \;\;\cm_{d}(p_A,p_B,p_C) \;\;, 
\eeq
where $\cm_f$ and $\cm_d$ are colour stripped amplitudes.
Owing to the Bose symmetry of $\ket{\cm_{ABC}}$ with respect to the three gluons,
the function $\cm_{f}(p_A,p_B,p_C)$ is antisymmetric under the exchange of two gluon momenta
(e.g., $p_A \leftrightarrow p_B$), while $\cm_{d}(p_A,p_B,p_C)$ has a symmetric dependence on
$p_A,p_B,p_C$.
Using Eqs.~(\ref{J2ggg})--(\ref{asymggg}), (\ref{triggg}) and (\ref{3gamp})
we obtain
\begin{align}
\label{m23g}
&\bra{\M_{ABC} (\{p_i\})} \;| \J(q_1,q_2,q_3) |^2 \;\ket{\M_{ABC} (\{p_i\})} 
= \left( \g\,\mu^\ep\right)^6 \,T_R \nn \\
&\;\;\;\; \times \Bigl\{ | \M_{ABC} (\{p_i\}) |^2 \; 
\Bigl[ \, F_{ABC}^{({\rm in.em.})}(q_1,q_2,q_3) 
+ W_{ABC}^{({\rm ch.sym.})}(q_1,q_2,q_3) \Bigr] \Bigr. \nn \\
&\;\;\;\;\;\;\; + \Bigl. C_A^2 d_A (N_c^2-1) 
\bigl[ \, \cm^\dagger_d(p_A,p_B,p_C)  \cm_f(p_A,p_B,p_C) + {\rm h.c.} \bigr]
E_{ABC}^{(tri)}(q_1,q_2,q_3)
\Bigr\} \;\;, \\
& \quad~~~~~~~~~~~~~~~~~~~~~~~~~~~ ~~~~~~~~~~~~~~~~~~~ ~~~~~~~~~~~~~~~~~~~~~~~~~~~\quad 
(\{A B C\}=\{ggg\}) \;, \nn
\end{align}
which is not simply proportional to $| \cm_{ABC}(\{p_i\}) |^2$ (unlike the corresponding result in Eq.~(\ref{sgqq})
for $\{ ABC \} = \{ gq{\bar q} \})$). 
The contribution to Eq.~(\ref{m23g}) that is proportional to 
$F_{ABC}^{({\rm in.em.})} + W_{ABC}^{({\rm ch.sym.})}$ is symmetric under the exchange
$q_2 \leftrightarrow q_3$ and, therefore, it does not lead to any charge asymmetry of the soft quark and antiquark. The function $E_{ABC}^{(tri)}(q_1,q_2,q_3)$ is instead 
antisymmetric under the exchange
$q_2 \leftrightarrow q_3$. Therefore, in contrast with the case of scattering amplitudes with two hard gluons (see Eq.~(\ref{BCgg})), the expression in Eq.~(\ref{m23g})
involves a charge-asymmetry contribution that is not vanishing,
provided the hard-scattering amplitude includes non-vanishing components $\cm_f$ and $\cm_d$
(i.e., $\ket{\cm_{ABC}}$ has no definite charge conjugation). This is the case, for instance,
of the amplitude for the decay process $Z \to ggg$ of the $Z$ boson
(see, e.g., Ref.~\cite{vanderBij:1988ac}).
We note that the functions $E_{ABC}^{(tri)}$ and $(\cm_d^\dagger \cm_f + {\rm h.c.})$ 
are separately antisymmetric under the exchange of two gluon momenta and, consequently, their
product is symmetric. Therefore, the right-hand side of Eq.~(\ref{m23g})
is fully symmetric under permutations of the three
hard gluons, as required by Bose symmetry.

\section{QED and mixed QCD$\times$QED interactions\label{s:qed}}

Our results of Sects.~\ref{s:gqqc} and \ref{s:gqqsc}
for soft $gq{\bar q}$ radiation at the tree level in QCD are generalized in this section to deal with soft emission through QED (photon) interactions and mixed QCD$\times$QED (gluon and photon)
interactions.

We consider a generic scattering amplitude $\M(\{q_\ell\}, \{p_i\})$ whose external soft {\it massless} particles are gauge bosons ($b$), fermions ($f$) and antifermions ($\bar f$).
The soft gauge bosons can be gluons ($b=g$) or photons ($b=\gamma$), and the soft massless fermions are quarks ($f=q$) and charged leptons ($f=\ell$). The external massless and massive hard partons in $\M(\{q_\ell\}, \{p_i\})$ are gluons, (anti)quarks and electrically charged particles, such as (anti)leptons and $W^\pm$ bosons. The amplitude $\M$ can also have external particles that carry no colour charge and no electric charge. 

We formally treat QCD, QED and mixed QCD$\times$QED interactions on equal footing. Therefore, the scattering amplitude $\M$ has a generalized perturbative (loop) expansion in powers of two unrenormalized couplings: the QCD coupling $\g$ and the QED coupling $\gq$ ($\gq^2/(4\pi)=\alpha$ is the fine structure constant at the unrenormalized level). In the soft limit
the amplitude $\M(\{q_\ell\}, \{p_i\})$ fulfils the factorization formula (\ref{1gfact}),
and the soft current $\J(q_1,\cdots,q_N)$ also has a loop expansion in powers of the two couplings $\g$ and $\gq$. In the following we only consider soft currents at the tree level
with respect to both couplings and, therefore, the $N$ parton current 
$\J(q_1,\cdots,q_N)$ includes all possible contributions that are proportional to 
$\g^{N-k} \gq^k$ with $0 \leq k \leq N$. The pure QCD and pure QED cases are recovered by setting $\gq=0$ and $\g=0$, respectively.

We first recall the known expressions of the soft currents for single-photon and
fermion-antifermion emission. The tree-level current $\bj_\gamma(q_1)$ for emission
of a single soft photon with momentum $q_1$ is
\begin{equation}
\label{J1ga}
  \bj_\gamma(q_1) = \gq \,\mu^\ep
\sum_{i} e_i \;\frac{p_i \pol(q_1)}{p_i q} \;\;,
\end{equation}
where $e_i$ is the electric charge (in units of the positron charge $\gq$) of the $i$-th
hard parton in $\M(\{p_i\})$. The conservation of the electric charge in $\M(\{p_i\})$
implies that $\sum_i e_i = 0$ (analogously to the colour conservation relation in 
Eq.~(\ref{colcons})). Note that $\bj_\gamma(q_1)$ is a $c$-number (more precisely, it is proportional to the unit matrix in colour space) since the photon carries no colour charge.
The square of the current in Eq.~(\ref{J1ga}) is 
\begin{equation}
\label{J2ga}
  |\bj_\gamma(q_1)|^2 = - \left(\gq \,\mu^\ep \right)^2 \sum_{i,k}  e_i\,e_k \;S_{ik}(q_1)
  \eqcs - \left(\gq \,\mu^\ep \right)^2 \frac12 \sum_{i\neq k}  e_i\,e_k \;w_{ik}(q_1)\;\;,
\end{equation}
where the momentum dependent functions $S_{ik}(q_1)$ and $w_{ik}(q_1)$ are given in 
Eqs.~(\ref{S1}) and (\ref{w1}).

The tree-level current $\bj_{f{\bar f}}(q_2,q_3)$ for emission of a soft-$f{\bar f}$ pair
is \cite{Catani:2021kcy}
\begin{align}
\label{J1ff}
  \bj_{f{\bar f}}(q_2,q_3) &= \delta_{fq} \;\bj(q_2,q_3)
  - \left( \gq \,\mu^\ep \right)^2  \,e_f \;{\bf \Delta}_f
  \, \sum_{i} e_i \;\frac{p_i  j(2,3)}{p_i  q_{23}} 
 \;,
\end{align}
where $q_2$ and $q_3$ are the momenta of the soft fermion $f$ and antifermion $\bar f$, respectively. The first contribution in the right-hand side of Eq.~(\ref{J1ff}) is the
QCD current $\bj(q_2,q_3)$ in Eq.~(\ref{Jqa})
(the Kronecker delta symbol $\delta_{fq}$ specifies that the current is not vanishing only
if $f=q$), and the second contribution is due to the photon mediated radiation of the 
$f{\bar f}$ pair. The fermionic current $j^\nu(2,3)$ is given in Eq.~(\ref{fercur}),
and $e_f$ is
the electric charge of the soft
fermion $f$.
The factor ${\bf \Delta}_f$ in the right-hand side of Eq.~(\ref{J1ff})
is a colour operator that depends on the type
of soft fermion $f$. If $f=\ell$, we simply have ${\bf \Delta}_f= 1$. If $f=q$, 
${\bf \Delta}_f$ is the projection operator onto the colour singlet
state of the $f{\bar f}$ pair, namely, by using the colour space notation of
Sect.~\ref{s:sfsc} we have
$\bra{\alpha_2,\alpha_3}  \,{\bf \Delta}_f = \delta_{\alpha_2\alpha_3}$, where
$\alpha_2$ and $\alpha_3$ are the colour indices of the soft quark and antiquark, respectively.

The square of the current $\bj_{f{\bar f}}(q_2,q_3)$ is \cite{Catani:2021kcy}
\begin{equation}
\label{J2ff}
|\bj_{f{\bar f}}(q_2,q_3)|^2 = \delta_{fq} \;|\bj(q_2,q_3)|^2
- \left(\gq \,\mu^\ep \right)^4 \,\left( \delta_{f\ell} + N_c \,\delta_{fq} \right) \,e_f^2\,\frac12 \sum_{i\neq k}  e_i\,e_k \;w_{ik}(q_2,q_3)\;\;,
\end{equation}
where $|\bj(q_2,q_3)|^2$ is the QCD squared current in Eq.~(\ref{j23q})
and the function $w_{ik}(q_2,q_3)$ is given in Eqs.~(\ref{I23}) and (\ref{w23}).
Similarly to its QCD part, the complete squared current $|\bj_{f{\bar f}}(q_2,q_3)|^2$
is charge symmetric with respect to the exchange $f \leftrightarrow {\bar f}$
(i.e., it is symmetric under $q_2 \leftrightarrow q_3$).
We note that the squared current result in Eq.~(\ref{J2ff}) does not include a mixed
QCD$\times$QED term proportional to $\g^2 \gq^2$, since such contribution vanishes.

In the following we present our results for soft $bf{\bar f}$ radiation at the tree level.
The boson $b$ has momentum $q_1$ and the fermion $f$ and antifermion $\bar f$ have momenta 
$q_2$ and $q_3$, respectively. Similarly to the results in Sect.~\ref{s:gqqc}, we express
the current $\bj_{bf\bar{f}}$ for soft $bf{\bar f}$ radiation in terms of an independent emission contribution and an irreducible correlation term $\bs\Gamma_{bf\bar f}$.

In the case of soft $gf{\bar f}$ radiation we obtain
\begin{equation}\label{gff}
  \bj_{gf\bar{f}}(q_1,q_2,q_3) = \sy{\bj(q_1)\,}{\bj_{f\bar f}(q_2,q_3)}
  + \bs\Gamma_{gf\bar f}(q_1,q_2,q_3) \;\;,
\end{equation}
where $\bj(q_1)$ is the QCD soft-gluon current in Eq.~(\ref{J1})
and $\bj_{f\bar f}(q_2,q_3)$ is the soft-$f{\bar f}$ current in Eq.~(\ref{J1ff}).
Introducing the explicit dependence on the colour index $a_1$ of the soft gluon,
the irreducible correlation $\bs\Gamma_{gf\bar f}$ is
\begin{equation}\label{corrgff}
  \bs\Gamma_{gf\bar f}^{a_1}(q_1,q_2,q_3) = \delta_{fq} \,
\Bigl[ \bs\Gamma^{a_1}(q_1,q_2,q_3) + \g\,\gq^2 \mu^{3\ep} e_f \sum_i e_i  \,
 \btq^{a_1} \,\gab_i(q_1,q_2,q_3) \Bigr] \;\;,
\end{equation}
where the term $\bs\Gamma^{a_1}(q_1,q_2,q_3)$ in the right-hand side is the QCD irreducible correlation in Eq.~(\ref{bfgamma}),
and the function $\gab_i(q_1,q_2,q_3)$ is given in Eq.~(\ref{psi}).
We note that the correlation $\bs\Gamma_{gf\bar f}$ is not vanishing only if $f=q$. The second term in the square bracket of Eq.~(\ref{corrgff}) is the mixed QCD$\times$QED correction to the QCD irreducible correlation for soft $gq\bar{q}$ emission.
We note that such QCD$\times$QED correlation term is proportional to 
$\gab_i(q_1,q_2,q_3)$ and, therefore, it has an abelian character.

The tree-level current for soft $\gamma f\bar{f}$ emission is
\begin{equation}\label{gaff}
  \bj_{\gamma f\bar{f}}(q_1,q_2,q_3) = {\bj_\gamma(q_1)\,}{\bj_{f\bar f}(q_2,q_3)}
  + \bs\Gamma_{\gamma f\bar f}(q_1,q_2,q_3) \;\;,
\end{equation}
where $\bj_\gamma(q_1)$ and $\bj_{f\bar f}(q_2,q_3)$ are the currents in 
Eqs.~(\ref{J1ga}) and (\ref{J1ff}).
We note that the independent emission contribution in Eq.~(\ref{gaff}) does not require
colour symmetrization, since $\bj_\gamma$ and $\bj_{f\bar f}$ commute in colour space.
The expression of the irreducible correlation component $\bs\Gamma_{\gamma f\bar f}$ is
\begin{equation}\label{corrgaff}
  \bs\Gamma_{\gamma f\bar f}(q_1,q_2,q_3) = \mu^{3\ep} e_f \sum_i \left[ \,
    \delta_{fq}\; \g^2\,\gq \; T_i^c \btq^c + \gq^3 \;
     e_f\, e_i \; \bs\Delta_f\right]  \,\gab_i(q_1,q_2,q_3)\;\;,
\end{equation}
where $\gab_i(q_1,q_2,q_3)$ is given in Eq.~(\ref{psi}).
The term proportional to $\gq^3$ in Eq.~(\ref{corrgaff}) is entirely due to QED
(photon) interactions. We note that even in an abelian gauge theory, like QED, the current
for soft $\gamma f\bar{f}$ emission includes an irreducible correlation component, which is due to soft-photon radiation in cascade from soft charged fermions. In contrast, we recall that the current for emission of $N$ soft photons factorizes in terms of $N$ independent emission contributions, with no additional irreducible correlations. The term proportional to
$\g^2 \gq$ in Eq.~(\ref{corrgaff}) is the irreducible correlation component that is due to mixed QCD$\times$QED interactions. Also this correlation component is controlled by the abelian function $\gab_i(q_1,q_2,q_3)$.

The squared current for soft $g f\bar{f}$ emission is computed by using the expressions in Eqs.~(\ref{gff}) and (\ref{corrgff}). We write the result as follows
\begin{equation}\label{J2gff}
  | \bj_{gf\bar{f}}(q_1,q_2,q_3) |^2
  = \sy{\,|\bj(q_1)|^2\,}{|\bj_{f\bar f}(q_2,q_3)|^2} + W_{gf\bar{f}}(q_1,q_2,q_3) \;\;,
\end{equation}
where $|\bj(q_1)|^2$ is the QCD squared current in Eq.~(\ref{j1q})
and $|\bj_{f\bar f}(q_2,q_3)|^2$ is the squared current in Eq.~(\ref{J2ff}). 
The irreducible correlation contribution $ W_{gf\bar{f}}$ is not vanishing only if the soft fermion is a quark ($f=q$), and we explicitly have
\begin{align}\label{wgff}
W_{gf\bar{f}}(q_1,q_2,q_3) &= \delta_{fq} \,\Bigl\{ W(q_1,q_2,q_3) \Bigr. \nn \\
&
\Bigl. - \,\g^4\,\gq^2\, \mu^{6\ep} \,2\,T_R\,e_f \sum_{i,k,m} \T_i\cdot \T_k \;e_m\;
  \bigr[ \cS_{ikm}(q_1,q_2,q_3)+(k \leftrightarrow m) \bigl] \Bigr. \nn \\
&\Bigl. - \,\g^2\,\gq^4\, \mu^{6\ep} C_F\,N_c\;e_f^2\sum_{i,k} e_i\, e_k\,
  \;\cS_{ik}^{(F)}(q_1,q_2,q_3) \Bigr\} \;\;,
\end{align}
where $W(q_1,q_2,q_3)$ is the QCD term in Eq.~(\ref{Wqcd}),
and the momentum dependent functions $\cS_{ikm}(q_1,q_2,q_3)$ and
$\cS_{ik}^{(F)}(q_1,q_2,q_3)$ are given in Eqs.~(\ref{Sikm}) and (\ref{SF}).
The right-hand side of Eq.~(\ref{wgff}) includes two types of mixed QCD$\times$QED contributions, which both have an abelian character. The contribution proportional to
$\g^4 \gq^2$ is controlled by the function $\cS_{ikm}(q_1,q_2,q_3)$ and, therefore,
it leads to charge asymmetry in the exchange of the soft quark and antiquark. In contrast,
the contribution proportional to $\g^2 \gq^4$ is charge symmetric, since it depends on the 
function $\cS_{ik}^{(F)}(q_1,q_2,q_3)$.

Using Eqs.~(\ref{gaff}) and (\ref{corrgaff}) the squared current for soft $\gamma f{\bar f}$ 
emission is
\begin{equation}\label{J2gaff}
  | \bj_{\gamma f\bar{f}}(q_1,q_2,q_3) |^2
  = {|\bj_\gamma(q_1)|^2\,}{|\bj_{f\bar f}(q_2,q_3)|^2} + W_{\gamma f\bar{f}}(q_1,q_2,q_3) \;\;,
\end{equation}
where $|\bj_\gamma(q_1)|^2$ is the soft-photon squared current in Eq.~(\ref{J2ga})
and $|\bj_{f\bar f}(q_2,q_3)|^2$ is given in Eq.~(\ref{J2ff}).
The irreducible correlation component $W_{\gamma f\bar{f}}$ has the following expression:
\begin{align}\label{wgaff}
W_{\gamma f\bar{f}}(q_1,q_2,q_3) =& - \,\g^4\,\gq^2\, \mu^{6\ep} \,\delta_{fq} \,T_R\,e_f
\nn \\
& \times \Bigl[ \;\sum_{i,k} e_f \,\T_i\cdot \T_k \;\cS_{ik}^{(F)}(q_1,q_2,q_3)
+ \sum_{i,k,m} 2 \,e_i \, \T_k\cdot \T_m \;\cS_{ikm}(q_1,q_2,q_3)
\Bigr]
\nn \\
& - \,\left(\gq \,\mu^\ep \right)^6 \,\left( \delta_{f\ell} + N_c \,\delta_{fq} \right) \,e_f^3 \nn \\
& \times \Bigl[ \;\sum_{i,k} e_f \,e_i\,e_k \;\cS_{ik}^{(F)}(q_1,q_2,q_3)
+ \sum_{i,k,m} 2 \,e_i \,e_k\,e_m \;\cS_{ikm}(q_1,q_2,q_3)
\Bigr]
 \;\;,
\end{align}
where the charge symmetric function $\cS_{ik}^{(F)}(q_1,q_2,q_3)$ is given in Eq.~(\ref{SF})
and the charge asymmetry function $\cS_{ikm}(q_1,q_2,q_3)$ is given in Eq.~(\ref{Sikm}).
The term proportional to $\gq^6$ in Eq.~(\ref{wgaff}) is entirely due to QED interactions.
The term proportional to $\g^4 \gq^2$ is due to mixed QCD$\times$QED interactions, and it is not vanishing only if the soft fermion is a quark. We note that both the QED and 
QCD$\times$QED contributions to $W_{\gamma f\bar{f}}$ involve charge symmetric and asymmetric effects. We also note that the contribution of ${\cal O}(\g^2 \gq^4)$
to $| \bj_{\gamma f\bar{f}}(q_1,q_2,q_3) |^2$ vanishes.

\section{Summary\label{s:conc}}

We have considered the radiation of a soft gluon and a soft $q{\bar q}$ pair
in QCD hard scattering. The scattering amplitude for soft $gq{\bar q}$ emission
in a generic hard-scattering process is singular, and the singular behaviour is controlled
in factorized form by a current $\J(q_1,q_2,q_3)$, which has a process-independent 
structure. 

We have evaluated the soft  $gq{\bar q}$  current $\J(q_1,q_2,q_3)$
at the tree level for a generic scattering amplitude with an arbitrary number and type
of external hard partons (gluons and massless and massive quarks and antiquarks).
The soft current acts in colour space, and it is written in terms of the colour charges and 
momenta of the external hard partons. We have expressed the current in terms of
two contributions: the contribution of `independent' (and colour symmetrized) emission
of the soft gluon and the soft $q{\bar q}$ pair, and an irreducible correlation contribution.
The irreducible correlation component of the current includes strictly non-abelian terms
(which are analogous to the non-abelian correlations for soft multi-gluon emission) and also
terms with an abelian character (analogous correlations appear for soft
photon-lepton-antilepton emission in QED).

We have computed the tree-level squared current $| \J(q_1,q_2,q_3) |^2$ of 
soft  $gq{\bar q}$ emission for squared amplitudes of generic multiparton hard-scattering
processes. We have checked that our result for $| \J(q_1,q_2,q_3) |^2$ numerically agrees
with the result obtained in Ref.~\cite{DelDuca:2022noh} in a fully independent way.
The irreducible correlation component of $| \J(q_1,q_2,q_3) |^2$ leads to two types
of colour interactions between the hard partons: colour dipole interactions
(which also appear in the independent emission component) and interactions of tripole type
that are proportional to the fully-symmetric tensor $d^{a b c}$. 
These tripole interactions are the real-emission counterpart of the analogous tripole interactions for soft-$q{\bar q}$ radiation at the one-loop level \cite{Catani:2021kcy}.
The tripole correlation
contributions to $| \J(q_1,q_2,q_3) |^2$ are antisymmetric under the exchange
$q_2 \leftrightarrow q_3$ of the momenta of the soft quark and antiquark and, therefore,
they produce charge asymmetry effects in the soft limit of the squared amplitudes.
We have explicitly considered the evaluation of $| \J(q_1,q_2,q_3) |^2$ for processes with two and three hard partons, and we have discussed the corresponding charge symmetric
and asymmetric contributions.

We have finally generalized our QCD study of soft $gq{\bar q}$ emission to the study
of QED and mixed QCD$\times$QED interactions in the context of soft gluon-fermion-antifermion and photon-fermion-antifermion radiation.   
In particular, we have noticed that soft photon-lepton-antilepton emission in QED received (abelian) irreducible correlation contributions due to soft-photon radiation in cascade
from soft charged leptons. Both QED and mixed QCD$\times$QED interactions lead to charge
asymmetry effects in the exchange of the soft fermion and antifermion.

\section*{Acknowledgments}

\includegraphics[width=2.2em,angle=90]{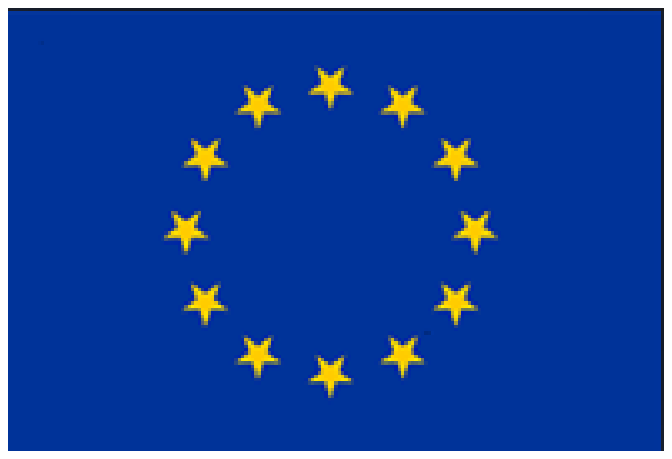}~
\begin{minipage}[b]{0.9\linewidth}
  This project has received funding from the European Union’s Horizon
  2020 research and innovation programme under grant agreement No
  824093. 
\end{minipage}
LC is supported by the Generalitat Valenciana (Spain) through the plan GenT program (CIDEGENT/2020/011) and his work is supported by the Spanish Government (Agencia Estatal de Investigación) and ERDF funds from European Commission (Grant no. PID2020-114473GB-I00 funded by MCIN/AEI/10.13039/501100011033).
\section*{Appendices}
\appendix

\section{Tripoles on two and three hard particles\label{a:tthp}}

In this appendix we present the action of the $d$-type colour tripoles in Eq.~(\ref{tripoli})
onto colour
singlet states of two or three QCD particles. The corresponding colour algebra was discussed in detais in Ref.~\cite{Catani:2021kcy}. In the following we limit ourselves to list
the explicit results \cite{Catani:2021kcy}.

As discussed in Sect.~\ref{s:e2h}, there are two possible colour-singlet QCD states
of two particles: $\ket{gg}$ and $\ket{q\qb}$. They are both eigenvectors of
all tripole operators. In fact, we have
\begin{align*}
  {\cal T}^{(d)}_{ikm}\ket{gg} &= 0 \;\;,\\
  {\cal T}^{(d)}_{ikm}\ket{q\bar{q}} &= \frac{(-1)^{I_{\bar q}}}{2}  d_A\,C_F\ket{q\bar{q}}
\;\;,
\end{align*}
where $I_{\bar q}$ is the number of indices ${i,k,m}$ corresponding to the antiquark.

As pointed out in Sect.~\ref{s:e3h}, we consider three colour-singlet
states formed with three QCD particles: $\ket{gq\qb}$,
$\ket{(ggg)_f}$ and $\ket{(ggg)_d}$.
The state $\ket{gq\qb}$ is an eigenvector of all colour tripoles. The
corresponding eigenvalues 
are summarized in the following table:\\

{\centering
\begin{tabular}{|c|c|}
  $\{i,k,m\}$ & $\ket{gq\bar q}$ \\
  \hline
  $ggg$ & 0 \\
  $ggq$ & $C_A d_A / 4$ \\
  $gg\bar q$ & $-C_A d_A / 4$ \\
  $gqq$ & $-C_A d_A / 4$ \\
  $g\bar q\bar q$ & $C_A d_A / 4$ \\
  $gq\bar q$ & $0$ \\
  $qqq$ & $C_F d_A / 2$ \\
  $\bar q\bar q\bar q$ & $-C_F d_A / 2$ \\
  $qq\bar q$ & $d_A(C_A - 2 C_F) / 4$ \\
  $q\bar q\bar q$ & $-d_A(C_A - 2 C_F) / 4$ \\
\end{tabular}\\
}

\vskip 3ex

\noindent and the remaining eigenvalues are obtained by exploiting their full symmetry under
permutations of the indices $i,k,m$.
In contrast, the colour tripoles swap the hard three-gluon states $\ket{(ggg)_f}$ and $\ket{(ggg)_d}$ in Eq.~(\ref{ABCbasis}), and we can write
\begin{equation}
  \tri{ikm}\ket{(ggg)_f} = \lambda^{(f)}_{ikm}\ket{(ggg)_d}\;,\qquad\qquad
  \tri{ikm}\ket{(ggg)_d} = \lambda^{(d)}_{ikm}\ket{(ggg)_f} \;.
\end{equation}
The values of the coefficients $\lambda^{(f)}_{ikm}$ and $\lambda^{(d)}_{ikm}$ for $i\leq k\leq m$ are
collected in the following table:

{\centering 
\begin{minipage}[c]{0.4\linewidth}
\end{minipage}
\begin{tabular}{|c|c|c|}
  $\{i,k,m\}$ & $\lambda^{(f)}$ & $\lambda^{(d)}$ \\
  \hline
  AAA & 0          & 0\\
  BBB & 0          & 0\\
  CCC & 0          & 0\\
  AAB & $-C_A^2/4$ & $-C_A d_A / 4$ \\
  ABB & $C_A^2$/4  & $ C_A d_A / 4$ \\
  AAC & $C_A^2$/4  & $ C_A d_A / 4$ \\
  ACC & $-C_A^2$/4  & $ - C_A d_A / 4$ \\
  BBC & $-C_A^2/4$ & $-C_A d_A / 4$\\
  BCC & $C_A^2/4$ & $C_A d_A / 4$\\
  ABC & 0          & 0 \\
\end{tabular}\\
}

\vskip 3ex

\noindent and the remaininfg coefficients are obtained by using their full symmetry under
permutations of the indices $i,k,m$.



\begin{thebibliography}{99}

\bibitem{Heinrich:2020ybq}
G.~Heinrich,
Phys. Rept. \textbf{922} (2021), 1-69
[arXiv:2009.00516 [hep-ph]].

\bibitem{Agarwal:2021ais}
N.~Agarwal, L.~Magnea, C.~Signorile-Signorile and A.~Tripathi,
[arXiv:2112.07099 [hep-ph]].


\bibitem{Becher:2014oda}
  T.~Becher, A.~Broggio and A.~Ferroglia,
  Lect.\ Notes Phys.\  {\bf 896} (2015) 1
  [arXiv:1410.1892 [hep-ph]].


\bibitem{Luisoni:2015xha}
  G.~Luisoni and S.~Marzani,
  J.\ Phys.\ G {\bf 42} (2015) no.10,  103101
  [arXiv:1505.04084 [hep-ph]].

\bibitem{Luo:2019szz}
M.~x.~Luo, T.~Z.~Yang, H.~X.~Zhu and Y.~J.~Zhu,
Phys. Rev. Lett. \textbf{124} (2020) no.9, 092001
[arXiv:1912.05778 [hep-ph]].

\bibitem{Ebert:2020yqt}
M.~A.~Ebert, B.~Mistlberger and G.~Vita,
JHEP \textbf{09} (2020), 146
[arXiv:2006.05329 [hep-ph]].

\bibitem{Luo:2020epw}
M.~x.~Luo, T.~Z.~Yang, H.~X.~Zhu and Y.~J.~Zhu,
JHEP \textbf{06} (2021), 115
[arXiv:2012.03256 [hep-ph]].

\bibitem{Ebert:2020qef}
M.~A.~Ebert, B.~Mistlberger and G.~Vita,
JHEP \textbf{07} (2021), 121
[arXiv:2012.07853 [hep-ph]].

\bibitem{Frixione:1995ms}
  S.~Frixione, Z.~Kunszt and A.~Signer,
  Nucl.\ Phys.\ B {\bf 467} (1996) 399
  [hep-ph/9512328].

\bibitem{csdip}
  S.~Catani and M.~H.~Seymour,
  Nucl.\ Phys.\ B {\bf 485} (1997) 291
   [Erratum:  Nucl.\ Phys.\ B {\bf 510} (1998) 503]
  [arXiv:hep-ph/9605323].

\bibitem{Frixione:1997np}
  S.~Frixione,
  Nucl.\ Phys.\ B {\bf 507} (1997) 295
  [hep-ph/9706545].

\bibitem{Catani:2002hc}
S.~Catani, S.~Dittmaier, M.~H.~Seymour and Z.~Trocsanyi,
Nucl. Phys. B \textbf{627} (2002), 189-265
[arXiv:hep-ph/0201036 [hep-ph]].


\bibitem{Campbell:1997hg}
J.~M.~Campbell and E.~W.~N.~Glover,
Nucl. Phys. B \textbf{527} (1998) 264-288
[arXiv:hep-ph/9710255 [hep-ph]].

\bibitem{Catani:1998nv}
S.~Catani and M.~Grazzini,
Phys. Lett. B \textbf{446} (1999) 143-152
[arXiv:hep-ph/9810389 [hep-ph]].

\bibitem{Bern:1998sc}
Z.~Bern, V.~Del Duca and C.~R.~Schmidt,
Phys. Lett. B \textbf{445} (1998) 168-177
[arXiv:hep-ph/9810409 [hep-ph]].

\bibitem{Kosower:1999rx}
D.~A.~Kosower and P.~Uwer,
Nucl. Phys. B \textbf{563} (1999) 477-505
[arXiv:hep-ph/9903515 [hep-ph]].

\bibitem{Bern:1999ry}
Z.~Bern, V.~Del Duca, W.~B.~Kilgore and C.~R.~Schmidt,
Phys. Rev. D \textbf{60} (1999) 116001
[arXiv:hep-ph/9903516 [hep-ph]].

\bibitem{Catani:1999ss}
  S.~Catani and M.~Grazzini,
  Nucl.\ Phys.\ B {\bf 570} (2000) 287
  [arXiv:hep-ph/9908523].

\bibitem{Catani:2000pi}
  S.~Catani and M.~Grazzini,
  Nucl.\ Phys.\ B {\bf 591} (2000) 435
  [arXiv:hep-ph/0007142].

\bibitem{Czakon:2011ve}
M.~Czakon,
Nucl. Phys. B \textbf{849} (2011) 250-295
[arXiv:1101.0642 [hep-ph]].

\bibitem{Bierenbaum:2011gg}
  I.~Bierenbaum, M.~Czakon and A.~Mitov,
  Nucl.\ Phys.\ B {\bf 856} (2012) 228
  [arXiv:1107.4384 [hep-ph]].

\bibitem{Czakon:2018iev}
M.~L.~Czakon and A.~Mitov,
[arXiv:1804.02069 [hep-ph]].

\bibitem{Catani:2011st}
S.~Catani, D.~de Florian and G.~Rodrigo,
JHEP \textbf{07} (2012) 026
[arXiv:1112.4405 [hep-ph]].

\bibitem{Sborlini:2013jba}
G.~F.~R.~Sborlini, D.~de Florian and G.~Rodrigo,
JHEP \textbf{01} (2014) 018
[arXiv:1310.6841 [hep-ph]].


\bibitem{Proceedings:2018jsb}
J.~R.~Andersen, J.~Bellm, J.~Bendavid, N.~Berger, D.~Bhatia, B.~Biedermann,
S.~Br\"auer, D.~Britzger, A.~G.~Buckley and R.~Camacho, \textit{et al.}
[arXiv:1803.07977 [hep-ph]].

\bibitem{Amoroso:2020lgh}
S.~Amoroso, P.~Azzurri, J.~Bendavid, E.~Bothmann, D.~Britzger, H.~Brooks,
A.~Buckley, M.~Calvetti, X.~Chen and M.~Chiesa, \textit{et al.}
[arXiv:2003.01700 [hep-ph]].

\bibitem{TorresBobadilla:2020ekr}
W.~J.~Torres Bobadilla, G.~F.~R.~Sborlini, P.~Banerjee, S.~Catani,
A.~L.~Cherchiglia, L.~Cieri, P.~K.~Dhani, F.~Driencourt-Mangin, T.~Engel and
G.~Ferrera, \textit{et al.}
Eur. Phys. J. C \textbf{81} (2021) no.3, 250
[arXiv:2012.02567 [hep-ph]].




\bibitem{DelDuca:1999iql}
V.~Del Duca, A.~Frizzo and F.~Maltoni,
Nucl. Phys. B \textbf{568} (2000) 211-262
[arXiv:hep-ph/9909464 [hep-ph]].

\bibitem{Birthwright:2005ak}
T.~G.~Birthwright, E.~W.~N.~Glover, V.~V.~Khoze and P.~Marquard,
JHEP \textbf{05} (2005) 013
[arXiv:hep-ph/0503063 [hep-ph]].

\bibitem{Birthwright:2005vi}
T.~G.~Birthwright, E.~W.~N.~Glover, V.~V.~Khoze and P.~Marquard,
JHEP \textbf{07} (2005) 068
[arXiv:hep-ph/0505219 [hep-ph]].

\bibitem{DelDuca:2019ggv}
V.~Del Duca, C.~Duhr, R.~Haindl, A.~Lazopoulos and M.~Michel,
JHEP \textbf{02} (2020) 189
[arXiv:1912.06425 [hep-ph]].

\bibitem{DelDuca:2020vst}
V.~Del Duca, C.~Duhr, R.~Haindl, A.~Lazopoulos and M.~Michel,
JHEP \textbf{10} (2020) 093
[arXiv:2007.05345 [hep-ph]].



\bibitem{Catani:2003vu}
S.~Catani, D.~de Florian and G.~Rodrigo,
Phys. Lett. B \textbf{586} (2004) 323-331
[arXiv:hep-ph/0312067 [hep-ph]].

\bibitem{Sborlini:2014mpa}
G.~F.~R.~Sborlini, D.~de Florian and G.~Rodrigo,
JHEP \textbf{10} (2014) 161
[arXiv:1408.4821 [hep-ph]].

\bibitem{Sborlini:2014kla}
G.~F.~R.~Sborlini, D.~de Florian and G.~Rodrigo,
JHEP \textbf{03} (2015) 021
[arXiv:1409.6137 [hep-ph]].

\bibitem{Badger:2015cxa}
S.~Badger, F.~Buciuni and T.~Peraro,
JHEP \textbf{09} (2015) 188
[arXiv:1507.05070 [hep-ph]].

\bibitem{Czakon:2022fqi}
M.~Czakon and S.~Sapeta,
JHEP \textbf{07} (2022), 052
[arXiv:2204.11801 [hep-ph]].



\bibitem{Bern:2004cz}
Z.~Bern, L.~J.~Dixon and D.~A.~Kosower,
JHEP \textbf{08} (2004)  012
[arXiv:hep-ph/0404293 [hep-ph]].

\bibitem{Badger:2004uk}
  S.~D.~Badger and E.~W.~N.~Glover,
  JHEP {\bf 07} (2004) 040
  [arXiv:hep-ph/0405236 [hep-ph]].

\bibitem{Duhr:2014nda}
C.~Duhr, T.~Gehrmann and M.~Jaquier,
JHEP \textbf{02} (2015) 077
[arXiv:1411.3587 [hep-ph]].



\bibitem{Li:2013lsa}
  Y.~Li and H.~X.~Zhu,
  JHEP {\bf 11} (2013) 080
  [arXiv:1309.4391 [hep-ph]].

\bibitem{Duhr:2013msa}
  C.~Duhr and T.~Gehrmann,
  Phys.\ Lett.\ B {\bf 727} (2013) 452
  [arXiv:1309.4393 [hep-ph]].

\bibitem{Dixon:2019lnw}
 L.~J.~Dixon, E.~Herrmann, K.~Yan and H.~X.~Zhu,
 JHEP \textbf{05} (2020) 135
 [arXiv:1912.09370 [hep-ph]].


\bibitem{Zhu:2020ftr}
Y.~J.~Zhu,
[arXiv:2009.08919 [hep-ph]].

\bibitem{Catani:2021kcy}
S.~Catani and L.~Cieri,
Eur. Phys. J. C \textbf{82} (2022) no.2, 97
[arXiv:2108.13309 [hep-ph]].


\bibitem{Catani:2019nqv}
  S.~Catani, D.~Colferai and A.~Torrini,
  JHEP {\bf 01} (2020) 118
  [arXiv:1908.01616 [hep-ph]].


\bibitem{DelDuca:2022noh}
V.~Del Duca, C.~Duhr, R.~Haindl and Z.~Liu,
[arXiv:2206.01584 [hep-ph]].


\bibitem{Feige:2014wja}
I.~Feige and M.~D.~Schwartz,
Phys. Rev. D \textbf{90} (2014) no.10, 105020
[arXiv:1403.6472 [hep-ph]].

\bibitem{Bassetto:1984ik}
  A.~Bassetto, M.~Ciafaloni and G.~Marchesini,
  Phys.\ Rept.\  {\bf 100} (1983) 201. 


\bibitem{Siegel:1979wq}
  W.~Siegel,
  Phys.\ Lett.\ B {\bf 84} (1979) 193.

\bibitem{Bern:1991aq}
  Z.~Bern and D.~A.~Kosower,
  Nucl.\ Phys.\ B {\bf 379} (1992) 451.

\bibitem{vanderBij:1988ac}
  J.~J.~van der Bij and E.~W.~N.~Glover,
  Nucl.\ Phys.\ B {\bf 313} (1989) 237.

\end{thebibliography}
\end{document}